\documentclass[12pt,preprint]{aastex}

\begin{document}

\title{Comparing the \ion{Ca}{2} H and K Emission Lines in Red Giant Stars}

\author{Jenna Ryon\footnote{REU student at McDonald Observatory}}
\affil{Harvey Mudd College, Dept. of Physics, Claremont, CA 91711}

\author{Matthew D. Shetrone}
\affil{University of Texas, McDonald Observatory, HC75 Box 1337-McD
    Fort Davis, TX, 79734}

\author{Graeme H. Smith}
\affil{UCO/Lick Observatory, University of California, Santa Cruz CA 95064}

\begin{abstract}
Measurements of the asymmetry of the emission peaks in the core of the 
\ion{Ca}{2} H line for 105 giant stars are reported. The asymmetry is 
quantified with the parameter $V/R$, defined as the ratio between the maximum 
number of counts in the blueward peak and the redward peak of the emission
profile. The \ion{Ca}{2} H and K emission lines probe the differential motion 
of certain chromospheric layers in the stellar atmosphere. Data on $V/R$ for 
the \ion{Ca}{2} K line are drawn from previous papers and compared to the 
analogous H line 
ratio, the H and K spectra being from the same sets of observations. 
It is found that the H line $(V/R)_H$ value is 
+0.04 larger, on average, than the equivalent K line ratio, however, the 
difference varies with $B-V$ color. Red giants cooler than 
$B-V=1.2$ are more likely to have $(V/R)_H > (V/R)_K$, whereas the 
opposite is true for giants hotter than $B-V=1.2$. The differences between the 
\ion{Ca}{2} H and K line asymmetries could be caused by the layers of 
chromospheric material from which these emission features arise moving with 
different velocities in an expanding outflow.
\end{abstract}

\section{Introduction}

The chromospheric emission component of the \ion{Ca}{2} K line has been studied
extensively in the optical spectra of giant stars (e.g., Wilson 1976, Blanco et
al. 1976, Stencel 1978, Kelch et al. 1978, Gray 1980, Middelkoop \& Zwaan 1981,
Middelkoop 1982, Rutten 1984). The \ion{Ca}{2} H line has not been used as 
often to probe the stellar atmosphere, though both lines exhibit emission in 
their cores that typically has a double-peaked profile. In the H line, the 
violet- and red-side emission peaks can be labeled H$_{2V}$ and H$_{2R}$ 
respectively. A re-absorption feature denoted H$_3$ is located at the core of 
the emission profile, and is produced higher in the chromosphere. The 
H$_{2V}$ and H$_{2R}$ peaks often do not have equal strength. To 
quantify this asymmetry, it is convenient to use the $V/R$ parameter of Wilson 
(1976) and Stencel (1978), which we define as the ratio of intensities in the 
H$_{2V}$ and H$_{2R}$ emission peaks. Asymmetric emission profiles with
$V/R$ values less than unity can be interpreted as evidence for an outflow 
having a differential velocity field in the chromospheric region, with the 
H$_3$ absorption feature being blueshifted with respect to the emission profile
(Stencel 1978). It is possible that the \ion{Ca}{2} H line probes a different 
level of the stellar chromosphere than the K line, depending upon the mechanism
for setting the relative populations in the upper states of these transitions 
(e.g., Avrett 1966; Linsky 1970). With the greater opacity of the K line 
leading to formation at higher altitudes, these emission lines can potentially 
be used to probe the onset of mass outflows within two different layers of the 
chromosphere.

In this paper we report the results from a spectroscopic program to measure the
$V/R$ parameter of the \ion{Ca}{2} H line for a sample of red giant stars. The 
asymmetry of the \ion{Ca}{2} H emission profile is compared to that of the K 
line. Our study uses spectra of 105 red giants observed during a ten-year 
period, and includes 98 stars previously investigated by Smith \& Shetrone 
(2000, 2004) and Shetrone et al. (2008).
The sample covers a range of stellar colors and luminosities in an effort 
to map out the behavior of the H emission as a function of location in the
H-R diagram. In addition, multiple observations were obtained for a number of
stars. Variability of the K emission line profile based on our data set has
been discussed by Shetrone et al. (2008). Although time variability is not the
intended subject of this paper, the individual observations listed in Table 1
allow an interested reader to judge the degree to which the H line asymmetry
can vary on a timescale of months. 

\section{Observations and Reduction}

High resolution spectra of the \ion{Ca}{2} H and K lines for a sample of 105 
field giants were obtained at McDonald Observatory. The spectra were acquired 
during several observing runs between October 1998 and June 2008 using either 
the Sandiford Cassegrain Echelle (CE) spectrometer 
on the 2.1-m Otto Struve Telescope or the 2d-coud\'{e} spectrometer on the
2.7-m Harlan J. Smith Telescope. For the 2.7-m we used a 1.2" slit
which yielded a 2.0 pixel resolution of $R=60,000$.
For the 2.1-m we used a 1.1" slit which yielded a 2.0 pixel resolution
of $R=60,000$.  This paper examines the 
\ion{Ca}{2} H line found in a different order of the same echelle spectra
for which K line observations have been reported upon by Smith \& Shetrone
(2000, 2004) and Shetrone et al. (2008). To this material we add more 
recent H and K line data from the spectra obtained in June 2008. 

The observing procedure and data reduction process for the spectra obtained
during observing runs from October 1998 through August 2007 are described in 
Smith \& Shetrone (2000, 2004) and Shetrone et al. (2008).
The most recent data were taken during the nights of 2008 June 22 and 23 
using the 2d-coud\'{e} spectrometer on the 2.7-m telescope. For this later 
run, the observing procedure was to take at least three exposures of 
each object with the maximum exposure time being 1200 s. 
Bias frames and flat field frames were taken at the end of each night. 
In the case of the 2.1-m program, since the spectrometer is a cassegrain
instrument, exposures of ThAr arcs were made before and after observation of 
each star. By contrast, the 2d-coud\'{e} spectrometer of the 2.7-m telescope 
is very stable, and for the observing runs with this instrument, ThAr
arcs were generally obtained only at the start and end of each night.
The spectra were bias corrected, flat-field corrected, and wavelength 
calibrated using routines within the IRAF \textbf{echelle} package. The 
individual spectra for each object taken on a given night were combined to 
remove cosmic rays. During the data reduction process, the number of observed 
electrons was preserved to keep the Poisson statistics valid for all coadded 
spectra. 
 
Using the IRAF \textbf{fxcor} and \textbf{dopcor} packages, the spectra were
shifted to the velocity frame of the Hinkle spectral Atlas of Arcturus (Hinkle
et al. 2000). The spectra were then divided by the Arcturus Atlas to obtain a
continuum residual, which was fit by a low order Legendre polynomial, excluding the core of the
\ion{Ca}{2} H line in the fitted sample. The combined spectra were divided by
the continuum fit to obtain normalized spectra.
This normalization procedure matches the procedures found in Smith \& Shetrone
(2004) and Shetrone et al. (2008).   The original spectrum, before continuum normalization, was also saved and referenced to determine
the Poisson noise.

\section{Results and Discussion}

\subsection{The Asymmetry of the \ion{Ca}{2} H$_2$ Emission}

Our sample is comprised of 105 giants that show double-peaked \ion{Ca}{2} 
H and K emission lines in their spectra. The stars observed are listed in 
Table 1 along with each date on which they were observed and the 
$(V/R)_H$ value for the H emission feature. The error (denoted 
$\epsilon_H$) on each $(V/R)_H$ value is calculated from the relation
$\epsilon_{H}^2 = \sigma_V^2/I_{H2R}^2 
     + I_{H2V}^2\sigma_R^2/I_{H2R}^4$, as
described in Shetrone et al. (2008). In this equation, the recorded photon 
counts in the H$_{2V}$ and H$_{2R}$ peaks of the unnormalized spectra 
are represented by $I_{H2V}$ and $I_{H2R}$, respectively. Taking the 
square root of $I_{H2V}$ and $I_{H2R}$ gives the Poisson errors in 
these counts, denoted $\sigma_V$ and $\sigma_R$, respectively. 

The possible effects of using a different template spectrum other than the
Arcturus atlas for the normalization procedure were explored using synthetic 
spectra. Model red 
giant spectra (normalized to their continua) 
were computed for effective temperatures of $T_{\rm eff} =$ 3800, 4130, 4540, 
and 5250 K, which span the range in temperature encompassed by the stars in 
our sample. Mismatches in spectral type associated with using these templates
in the normalization of the observed spectra were found to introduce possible 
systematic effects in derived $(V/R)$ values on the level of $\approx 0.005$, 
which is relatively small compared to uncertainties due to photon-statistics
(see Table 1). For the random errors associated with the normalization process,
which vary according to where the fitting points are set outside the H and K 
lines, the errors are typically 0.007 on $(V/R)_K$ and 0.003 on $(V/R)_H$, which 
are again considerably smaller than Poisson noise in the photon counts. 
While the exact template used for normalization is not critical, it is 
important that the normalization be done. A test of the $(V/R)$ values
in the unnormalized spectra revealed differences of --0.02 to --0.05 
for the $(V/R)_H$ depending upon the color of the star.   

As noted, we have simultaneous spectra of the \ion{Ca}{2} K emission
line for each H-line observation listed in Table 1, these two features falling
in different echelle orders of the same CCD exposure. 
The $(V/R)_K$ value of the \ion{Ca}{2} K emission feature for each
observation, and its error $\epsilon_K$, are also reported in Table 1.
These values were taken from Smith \& Shetrone (2004) and Shetrone et al. 
(2008), with the exception of the June 2008 measurements. Several 
$(V/R)_{K}$ values were taken from Smith \& Shetrone (2000) and adjusted 
with a zero point offset of $+0.026$ to correct for differences between 
normalization procedures, as described in Shetrone et al. (2008).
The Smith \& Shetrone (2000, 2004) papers did not include errors in the 
$(V/R)_K$ data. To estimate these we first determined an average difference 
between the errors in $(V/R)_H$ and $(V/R)_K$ from the Shetrone et al. 
(2008) data and the H analysis done here. This average difference, 
$\epsilon_H - \epsilon_K$, is 0.005. This offset was applied to the 
$\epsilon_H$ values derived from our analysis of the older \ion{Ca}{2} H line 
spectra to estimate corresponding values of $\epsilon_K$.
These new $\epsilon_K$ error estimates, the $\epsilon_K$ errors from Shetrone 
et al. (2008), and the $\epsilon_H$ errors from this analysis are listed in 
Table 1. 

Photometry in the Johnson $BV$ system was taken from the General Catalogue of 
Photometric Data (GCPD; Mermilliod et al. 1997), accessed from an online 
catalog\footnote{Available at http://obswww.unige.ch/gcpd/gcpd.html.}. The 
$B-V$ color of each star, as well as the absolute visual magnitude derived 
from Hipparcos parallaxes (assuming zero interstellar absorption), are 
listed in Table 2. The parallaxes were tabulated in the earlier papers of Smith
\& Shetrone (2002, 2004). The distance range among most of the stars in our 
sample is 20-200 pc, with 9 stars of known parallax being beyond 200 pc.
Also listed in Table 2 is the heliocentric radial velocity $v_r$ as obtained 
from the SIMBAD database and the metallicity from McWilliam (1990) of each
star. These properties are all consistent with our sample being Population I 
giants. 

In Figure 1 the asymmetry $(V/R)_H$ of the \ion{Ca}{2} H line is plotted 
versus the $B-V$ color of each star. In cases of stars observed multiple times,
each observation is plotted separately, and consequently this diagram contains 
the effects of any time variability in the H line asymmetry. With the exception 
of one observation of HD 203387, only stars with $B-V > 1.2$ exhibit 
red-dominant asymmetry where $(V/R)_H < 1.0$. The most extreme red asymmetries 
where $(V/R)_H < 0.7$ are only seen in stars with $B-V > 1.5$. Figure 1 can be 
compared to Figure 2 from Shetrone et al. (2008), which gives an analogous plot
for the asymmetry parameter of the \ion{Ca}{2} K line. Both plots illustrate 
two groupings of data points spanning similar asymmetry and color ranges: one 
in the region $1.0 < (V/R) < 1.3$ and $0.85 < (B-V) < 1.05$, and the other with
$0.7 < (V/R) < 1.4$ and $1.2 < (B-V) < 1.45$. However, data for giants
redder than $B-V = 1.45$ are unique to Figure 1, which represents a more
diverse sample of stars than that in Shetrone et al. (2008). 

\subsection{The Issue of Interstellar Absorption}

Stellar H and K emission line profiles can be affected by \ion{Ca}{2} absorption 
in the interstellar medium (ISM). Although we do not have spectra of the 
interstellar \ion{Ca}{2} lines in the directions of our program stars, several 
types of evidence indicate that the asymmetry trends seen in the
$(V/R)$  data are not attributable to systematic interstellar corruption of the
stellar \ion{Ca}{2} lines.
 
If interstellar absorption was dominating the observed $(V/R)$ values then the 
sense of the asymmetry should correlate with the velocity of each star relative
to the ISM (as in Figure 5 of B\"{o}hm-Vitense 1981 for the \ion{Mg}{2} $k_2$ 
emission). Several diagrams were prepared to test whether a similar correlation
is present in the \ion{Ca}{2} data. Plots were made of the average $(V/R)_K$ 
and average ($V/R)_H$ values of each star versus heliocentric radial velocity 
$v_r$; no trend is seen (Figure 2). Furthermore, a version of Figure 1 was 
made using different symbols to depict stars in different radial velocity 
ranges: $v_r < -20$ km s$^{-1}$, $-20 < v_r < +20$ km s$^{-1}$, and  $v_r > 20$
km s$^{-1}$. There is no evidence that stars with positive or negative $v_r$
occupy systematically different regions of the diagram. We conclude 
that the \ion{Ca}{2} emission lines are more resilient to ISM corruption than 
found by B\"{o}hm-Vitense (1981) for the \ion{Mg}{2} $h$ and $k$ emission. 

As a second test, the ISM absorption $A_V$ towards most program stars was 
calculated using the numerical recipe of Hakkila et al. (1997), which takes 
into account both the distance to each star and the Galactocentric coordinates
of each star.
Many of the stars from Table 1 are within 75 pc of the Sun, and within a 
local bubble inside of which ISM extinction is considered to be small or 
negligible (e.g., Sfeir et al. 1999; Luck \& Heiter 2007). The equivalent width
of the interstellar \ion{Ca}{2} K line as seen against early-type stars within 
100 pc of the Sun tends to be less than 10 m\AA\ (e.g., Vallerga et al. 1993), 
although exceptions occur. The values of $A_V$ derived according to the Hakkila
et al. (1997) model, modified such that $A_V$ is set to zero within 75 pc, are
listed in Table 2. For the majority of program stars, the estimated
$A_V < 0.1$ mag. Furthermore, those stars with the largest values of $A_V$ do
not occupy systematically discordant positions in the H and K line $(V/R)$
versus $(B-V)$ diagrams. In summary, considering
both the radial velocities of the program stars, and the estimated interstellar
extinction towards them, we see no systematics of the type that would be
expected if ISM corruption of the \ion{Ca}{2} H and K emission profiles was 
governing the observed $(V/R)$ values.

\subsection{Differences Between the H and K Line Asymmetries}

Despite the similarities between Figure 1 of this paper and Figure 2 of
Shetrone et al. (2008), there are nonetheless differences between
$(V/R)_H$ and $(V/R)_K$ for many stars. Some illustrative examples 
of asymmetry differences are shown in Figures 3 and 4. The first of these
shows spectra of HD 156283 obtained on 22 June 2008 using the 2.7-m telescope, 
wherein the $(V/R)_K$ ratio is $0.99$ and $(V/R)_H$ is $1.07$. At this time, 
although the K emission line was nearly symmetrical, the H line showed a 
blue-enhanced asymmetry in which the violet H$_{2V}$ peak was stronger than 
the red peak. Figure 4  shows a spectrum of HD 21552 obtained on 3 October 2001
using the 2.1-m  telescope. The \ion{Ca}{2} $(V/R)_K$ ratio for this 
spectrum is $0.94$ while $(V/R)_H = 1.01$. In this case, although the H 
line is nearly symmetric, the K emission profile is stronger in the red peak.

The difference between the \ion{Ca}{2} H and K line $V/R$ values was 
calculated for every observation of each star in the sample. The resulting
set of difference values is plotted versus the H line $(V/R)_H$ in 
Figure 5; each observation of each star is plotted, 
so some stars are represented by multiple points. The error bars were 
calculated by summing $\epsilon_H$ and $\epsilon_K$ in quadrature. 
The value of $(V/R)_H$ is larger than $(V/R)_K$ 
by an average offset of $+0.039$, which is shown as a dashed line in
Figure 5. However, it can be seen that there are trends within this diagram, 
such that a single average offset is not a good fit to the entire data set.

The difference between the H and K line $V/R$ values are shown in Figure 6 
versus $B-V$ color. As with Figure 5, asymmetry values from individual spectra 
are plotted, so that some stars are depicted with multiple data points. The 
linear best fit to the data is plotted as a dashed line, which has a slope of 
$0.162 \pm 0.017$. However, this fit does not seem to be the best 
representation of the data. In order to smooth over possible intrinsic 
variability, an average of the $(V/R)_H - (V/R)_K$ values was calculated for 
each star. These averages are shown plotted versus $B-V$ in Figure 7, so that 
unlike Figure 6, each star is depicted by just one data point. In order to 
further smooth the data to look for trends, a weighted mean was made of every 
five points in Figure 7 binned according to $B-V$. A continuous line shows 
the locus of these binned averages. Rather than a linear relation, 
these binned data exhibit a quasi-step function with a reasonably abrupt 
change at $(B-V) \sim 1.25$. The two horizontal lines in the figure show
the average value of the asymmetry difference among giants with colors
on either side of this ``divide.''

The color-magnitude diagram for our sample of red giants is plotted in Figure 
8. The data points are coded according to the significance level of the 
$(V/R)_H - (V/R)_K$ values; open circles and open squares
represent stars for which the average asymmetries differences are 
$\langle (V/R)_H - (V/R)_K \rangle > 1\sigma$ and 
$\langle (V/R)_H - (V/R)_K \rangle < 1\sigma$, respectively, while 
filled triangles represent stars for which
$\vert \langle (V/R)_H - (V/R)_{\rm K} \rangle \vert < 1\sigma$.
Red giants for which $(V/R)$ is notably greater in the H line than the K 
line (open circles) tend to have colors of $(B-V) > 1.2$, while stars for 
which the reverse is the case (open boxes) are mainly of $(B-V) < 1.2$, 
although they do extend as red as $(B-V) = 1.4$. Giants with similar H and K 
line asymmetries (filled triangles) are evenly spread across the entire $B-V$ 
range covered by the data sample. Thus, stars with similar asymmetry 
differences tend to group in specific regions of the color-magnitude diagram, 
albeit with some overlap, particularly in the color range $1.2 < (B-V) < 1.4$. 
This is a further reflection of the trend seen in Figure 6.

\subsection{Photospheric Effects}

The \ion{Ca}{2} H and K line chromospheric emission components are 
superimposed on a photospheric absorption spectrum that varies with both
wavelength and effective temperature. Could the differences between 
$(V/R)_{\rm H}$ and $(V/R)_{\rm K}$ be due to differences in the background
photospheric spectrum in the vicinity of the H and K lines? We have 
investigated this question through the use of synthetic photospheric
spectra calculated using the 2007 version of the program MOOG (Sneden 1973).
These spectra were generated for a model atmosphere with $T_{\rm eff} = 4250$ 
K, $\log g = 1.75$, and [Fe/H] = --0.1, a combination of parameters that is 
typical of the red giants in our sample. Synthetic photospheric spectra were 
generated for three linelists: one with all of the photosphere lines, a second 
list from which the H and K lines had been removed, and a third that
contained only the \ion{Ca}{2} H and K lines. To simulate a symmetric 
double-peaked H or K emission feature, two equal-height Gaussian profiles were 
combined having separations in wavelength that are typical of the observed 
spectra. In accordance with the optically thin 
case, the relative strengths of these H and K emission profiles were set to be 
1:2, with the peak flux in the K emission being 15\% that of the continuum 
level of the photospheric spectrum on which it is superimposed.
Mathematically, the flux of these artificial peaks was simply added to the 
synthetic photospheric spectrum computed with the full line list.

A decomposition of the resultant model spectrum is presented in the various 
panels of Figure 9, which show from top to bottom: the H and K photospheric 
lines alone, all photospheric lines other than H and K, the complete 
photospheric spectrum, the artificial H and K emission profiles,
and the full photospheric spectrum plus the artificial emission lines.
There are numerous weak and moderate strength absorption lines throughout
the wavelength range $3925 < \lambda < 3980$ \AA, but once they get 
superimposed on the very broad and deep photospheric profiles of the H and K
lines, the amplitudes of these weaker lines are greatly reduced. Since the 
total photospheric flux in the cores of these lines is relatively small, 
the effect of the photospheric background on the chromospheric emission
profiles is very modest. In Figure 9, the asymmetry induced by this effect
is barely discernible. 

To complement the simulation, photospheric
models were also computed for a range of effective temperatures. As
with Figure 9, simulated H and K emission lines, both with $(V/R) = 1$ 
and a K line twice as strong as the H line, were added to the set of 
synthetic photosphere spectra. From these superimposed spectra the values
of $(V/R)_K$, $(V/R)_H$, and $(V/R)_H - (V/R)_K$ were calculated.
In addition to simulations with symmetric emission profiles, calculations 
were also done for input values of $(V/R)_{H,K} = 1.2$ and 0.6, as well 
as a range of wavelength separations between the two maxima in each 
bimodal profile. In all cases the input H and K emission profiles had 
identical asymmetry, such that initially $(V/R)_H -(V/R)_K = 0$.
The value of this difference after superimposing the model photosphere 
spectrum illustrates the extent to which photospheric corruption might 
be affecting the observed asymmetry differences.
The resultant values of $(V/R)_H - (V/R)_K$ are plotted versus $B-V$ in
Figure 10. Modest systematic variations are seen
as a function of $(B-V)$, which are largely driven by photospheric
corruption of the K line rather than the H line. However, the magnitude 
of these asymmetry differences is much smaller (by a factor of 5) than
the star-to-star differences, and their trend with $(B-V)$, than we actually
observe. Although this modeling approach is no substitute for a self-consistent
treatment of radiative transfer through a combined model photosphere plus
chromosphere, it does suggest that the $(V/R)$ asymmetry variations
among red giants are not an artifact of photospheric differences, but reflect 
physical differences in the stellar chromospheres.

In computing the synthetic spectra it was assumed that there is no 
systematic velocity shift between the simulated emission profiles and the 
photospheric spectrum. As noted above, in the data reduction procedure all of 
the observed spectra were shifted so that the photospheric lines are at the 
rest wavelengths as defined by the Arcturus atlas. The average central 
wavelengths of the H and K emission profiles were measured in this rest 
frame for the 8 stars observed at the 2.7-m telescope in 2008. There was no 
systematic displacement of the emission line centers from the photospheric 
reference frame. The average deviation was found to be $-0.02$ km s$^{-1}$ 
with a standard deviation of 1.0 km s$^{-1}$ (1 pixel = 2.4 km s$^{-1}$),
and there is no trend with $(B-V)$ color.

\section{Conclusions}

Values of the asymmetry parameter $(V/R)_H$ for the \ion{Ca}{2} H 
chromospheric emission line have been measured for a large sample of red giant 
stars. The results are presented in Table 1, plotted as a function of $B-V$ in 
Figure 1, and in a series of other diagrams are compared
to $V/R$ values of the \ion{Ca}{2} K line based on the same spectroscopic 
data set (see Shetrone et al. 2008). Many of the spectra exhibit 
opposite asymmetries between the H and K line emission peaks. The 
H line $(V/R)_H$ value is larger than the $(V/R)_K$ ratio for the majority of 
stars in our sample, however, this difference shows a trend with 
$B-V$ color, being largest among giants with $(B-V) > 1.25$.

The observations can be interpreted within a framework in which the H and K 
emission lines are formed at different levels in a stellar chromosphere, 
thereby allowing for a comparison between the motions of these layers. 
Asymmetry in either the H or K line, i.e., $V/R$ different from unity,
is generally attributed to Doppler shifts resulting from a differential
velocity field in the chromosphere. In this Doppler picture, gas that produces 
the H$_{2V}$ and H$_{2R}$ peaks is taken to be moving with a different velocity
to the gas that contributes to the central H$_3$ reversal. This causes the
H$_3$ reversal to be redshifted or blueshifted with respect to the H$_2$ peaks,
resulting in an emission line asymmetry. Such a scenario has been invoked,
for example, to interpret H and K line asymmetry differences among dwarf 
stars (Rauscher \& Marcy 2006). 

On the Sun the $V/R$ values for the H emission line follow a slightly
skewed distribution with a maximum at unity, a minimum of 
$(V/R)_H \approx 0.5$, and an extended tail to values greater than $1.7$ (Cram 
\& Dam\'{e} 1983). There is only a weak correlation between $(V/R)_{\rm H}$ 
and the strength of the H emission, from which Cram \& Dam\'{e} infer that 
``the $V/R$ ratio of the integrated solar H line is not due specifically to 
bright (e.g., network) elements''. Furthermore,
the H$_3$ reversal is found to be consistently displaced away from whichever
of the H$_{2V}$ and H$_{2R}$ peaks is brightest, which indicates that 
Doppler velocity shifts of the H$_3$ reversal are an important factor governing
the asymmetry of the H emission profile (Cram \& Dam\'{e} 1983). Enhanced 
emission in the H$_{2V}$ peak appears to be associated with shock heating
(e.g., Beck et al. 2008). In regions of the quiet Sun, the H$_{2V}$ and H$_{2R}$
emission have even been observed to disappear, indicative of cool elements 
within cells of the chromospheric network (Rezaei et al. 2008).

Among the majority of dwarf stars with spectral types in the range
K7-M5, Rauscher \& Marcy (2006) show that $V/R > 1$ in both the H and K lines,
with a preference for $(V/R)_ K > (V/R)_H$. In addition, the
K$_3$ reversal tends to have a systematically greater redshift than the 
H$_3$ feature. It is worth quoting the conclusions of Rauscher \& Marcy on this
point: ``While the unequal degeneracies in the upper levels producing these
lines cause somewhat differing optical depths, these data emphasize that the 
H and K lines are produced in slightly different regions of the chromosphere
and are therefore governed by slightly different line transfer and
kinematics.'' The redshift of the central reversals is attributed by them
to micro-flare activity that heats chromospheric gas and propels it outwards.
As it moves upward it cools and decelerates, resulting in the gas that
contributes to the central reversal being redshifted relative to the
faster moving, hotter gas below it that produces the emission peaks. The
velocity differential between these regions of gas is on the order of several
tenths of a km s$^{-1}$. The greater redshift of the K$_3$ reversal
compared to the H$_3$ reversal is consistent with the former arising from 
slightly higher altitudes in the chromosphere, where
deceleration has proceeded to a greater degree.

Clues to the systematics of the chromospheric motions at work within our
sample of red giants can be found in Figures 11 and 12. Average values of 
$(V/R)_H$ and $(V/R)_K$ were calculated for each star using the data in Table 
1. The main panel of Figure 11 shows these average asymmetries plotted versus 
one another. In Figure 12 both $\langle (V/R)_H \rangle$ and 
$\langle (V/R)_K \rangle$ are plotted versus the average value of the 
asymmetry difference $\langle (V/R)_H - (V/R)_K \rangle$. This diagram is 
analogous to Figure 5, except that all data for a given star are averaged, so 
that each giant is represented by a single point.  These figures map out
the relative behavior of the H and K emission line profiles.

Among giants with the largest values of $(V/R)_H$ ($> 1.1$) this 
asymmetry parameter is nearly equal in value to $(V/R)_K$; such stars 
exhibit a blue-dominant asymmetry in both the H and K emission profiles, and 
scatter closely around the locus $(V/R)_H = (V/R)_K$ which is 
shown by the straight line in Figure 11. Among such stars it seems that the H 
and K lines are evincing similar velocity fields and that there is no evidence 
of systematic outflow in the chromosphere. 

Turning to giants with $1.0 < (V/R)_H < 1.1$, there is a noticeably larger 
range of K line asymmetries and one starts to see a systematic shift of the K 
line profiles to red-dominant asymmetries, i.e., to $(V/R)_K < 1.0$. A 
significant fraction of giants with $1.0 < (V/R)_H < 1.1$ also have 
$(V/R)_K < 1.0$, and so exhibit opposite H and K emission profile asymmetries. 
Among such stars we infer that an accelerating outflow has become established 
within a region of the chromosphere that spans altitudes at which the K$_2$ and
K$_3$ features are formed. We interpret the later feature as being formed 
at higher levels, in regions of faster outflow, than K$_2$. This causes the 
central absorption reversal to be blueshifted with respect to the emission 
profile, thereby giving a characteristic asymmetry of $(V/R)_K < 1.0$. However,
the outflow does not yet extend to the deeper altitudes where the H$_3$ feature
is formed. It is therefore a change in the K line asymmetry that drives the 
non-zero values of $\langle (V/R)_H - (V/R)_K \rangle$ among giants having
$1.0 < (V/R)_H < 1.1$. As a result, many of these stars fall below the straight
line in Figure 11. This phenomenon is reflected by a plateau of approximately 
constant $(V/R)_H \sim 1.1 \pm 0.05$ in the upper panel of Figure 12, whereas
the same stars fall along an inclined locus in the lower panel.

Eventually the accelerating outflow extends downwards in altitude to regions 
where the H$_3$ feature is formed, resulting in $(V/R)_H$ now decreasing to a 
value below unity. This brings about a maximum in the mean difference 
$\langle (V/R)_H - (V/R)_K \rangle$ of $\sim 0.2$. Both $(V/R)_H$ and $(V/R)_K$
are now less than unity and we infer that an accelerating wind now extends 
throughout the chromospheric regions where the K$_3$, K$_2$, H$_3$, and 
possibly the H$_2$ features originate. As the value of $(V/R)_H$ drops below 
$0.8$ the difference in $V/R$ between the H and K lines diminishes, and for 
stars with $(V/R)_H \sim 0.4$-0.5 the asymmetry parameters for the two lines 
become approximately equal. Such giants may be those with the largest velocity 
gradients in the chromosphere. They have returned to the vicinity of the
straight line plotted in Figure 11.

Figure 1 shows that the systematic trends outlined in the previous   
paragraphs correspond approximately to a sequence in $B-V$ color. Thus it may 
be that as stars evolve up the red giant branch they travel along an 
evolutionary sequence in Figures 11 and 12, with winds first originating at 
high altitudes in the chromosphere, and then becoming more established and 
extending to lower altitudes. 

However, it is also known that $V/R$ for the H and K lines 
can be time variable (Shetrone et al. 2008), 
and this is likely to impose scatter on any long-term evolutionary sequence.
An excellent example is the star Arcturus, which both Gray (1980) and 
Qin \& Li (1987) find to be variable over timescales of hours to months.
This indicates that red giant atmospheres 
are likely subject to both short and long term dynamical modification. Changes 
on a timescale of hours were interpreted by Qin \& Li (1987) as being due to 
a ``chromospheric eruption'' that may have ejected mass rapidly through the 
chromosphere and possibly to escape from the star. 

In the solar chromosphere significant differences in the ratio of H and K line 
source functions can occur over a range in height of approximately 500-700 km 
(Linsky 1970). Comparing to a G giant of solar effective temperature 
and mass, and radius $R = 40 R_{\odot}$, the chromospheric scale height might 
be expected to vary roughly as the inverse surface gravity, i.e., as $R^2$. 
Thus, variations in the ratio of the H and K line source functions might extend
over an altitude range of $\approx 800,000$ km. A disturbance moving at
5 km s$^{-1}$ could traverse such a region in about 44 hours. Although
we do have multiple observations of many stars, the typical time difference
between consecutive spectra is months or years rather than days.
Thus, in general, we don't have fine enough temporal coverage to detect 
possible systematic variations in H or K line asymmetries due to
passage of a short-term ``event'' through the chromosphere.
In the case of HD 27697, the two spectra obtained in 1998 on consecutive
nights reveal no evidence for any significant differences in
$(V/R)$ for either the H$_2$ or K$_2$ features. Similarly, spectra of HD 82635
taken 18 nights apart in 2007 show no significant evidence of variability.
In 2006, the data show that both the H$_2$ and K$_2$ $(V/R)$ values for HD 222404 
differed significantly between 1 September and 4 October, but our observations 
were a month, rather than several days, apart.

It should be noted, however, that the H and K line observations of Arcturus 
by Qin \& Li (1987) may depict a very energetic version of a short-timescale
phenomenon. The asymmetry indices measured by them (using their notation where
$r$ refers to the intensity in an emission line peak) vary 
within the range $0.87 < r({\rm H}_{2V})/r({\rm H}_{2r}) < 1.21$ and 
$0.79 < r({\rm K}_{2V})/r({\rm K}_{2r}) < 0.93$ over a period of
4 hours, and all their observations show 
$r({\rm H}_{2V})/r({\rm H}_{2r}) > r({\rm K}_{2V})/r({\rm K}_{2r})$.
Furthermore, in 4 of their 6 spectra $r({\rm K}_{2V})/r({\rm K}_{2r}) < 1.0$ 
while $r({\rm H}_{2V})/r({\rm H}_{2r}) > 1.0$, i.e., the H and K emission 
lines evince opposite asymmetries. On the basis of this example, it follows 
that short-term variability could indeed contribute to the star-to-star scatter
in asymmetries seen in the McDonald spectra. Such effects may be superimposed 
on a much longer-term evolutionary sequence that produces the trends with 
$(B-V)$ color seen in Figures 11 and 12, and which may be associated with 
the onset of pulsation and/or the driving of a large scale wind.

\acknowledgments

This project was completed as part of the McDonald Observatory REU program 
and was supported under NSF AST-0649128.  This research has made use of
the Simbad database, operated at CDS, Strasbourg, France.   We would like
to thank E. Luck for providing the extinction code, hakkila,  
used in this work.

\clearpage
\begin{deluxetable}{lrrrrr}
\tablecaption{ Individual measurements }
\tablehead{
\colhead{Star} &
\colhead{Date } &
\colhead{$(V/R)_H$} &
\colhead{error} &
\colhead{$(V/R)_K$} &
\colhead{error} \\
\colhead{(HD)} &
\colhead{} &
\colhead{(H)} &
\colhead{($\epsilon_H$)} &
\colhead{(K)} &
\colhead{($\epsilon_K$)}}
\startdata
1013  &  2001-10-05  &  0.79  &  0.03  &  0.70  &  0.03     \\
\hline
1227  &  1998-11-28  &  1.15  &  0.07  &  1.13  &  0.06     \\
\hline
1522  &  2006-10-03  &  1.14  &  0.05  &  1.03  &  0.05     \\
\hline
3627  &  1998-10-25  &  1.16  &  0.04  &  1.10  &  0.04     \\
      &  2001-10-08  &  1.08  &  0.02  &  1.09  &  0.03     \\
\hline
4128  &  2006-08-31  &  1.07  &  0.03  &  1.12  &  0.03     \\
      &  2006-10-04  &  1.04  &  0.02  &  1.07  &  0.03     \\
\hline
4482  &  1998-10-25  &  1.11  &  0.03  &  1.10  &  0.03     \\
\hline
4730  &  2001-10-08  &  1.06  &  0.06  &  1.06  &  0.05     \\
\hline
5516  &  1998-11-30  &  1.10  &  0.04  &  1.22  &  0.03     \\
\hline
6953  &  2001-10-05  &  1.14  &  0.05  &  1.08  &  0.04     \\
\hline
7318  &  1998-11-26  &  1.16  &  0.06  &  1.14  &  0.05     \\
\hline
9138  &  2001-10-04  &  1.03  &  0.07  &  0.92  &  0.04     \\
      &  2006-10-04  &  0.74  &  0.05  &  0.97  &  0.09     \\
\hline
9774  &  1998-11-29  &  1.16  &  0.06  &  1.18  &  0.06     \\
\hline
11559  &  1998-11-28  &  1.13  &  0.05  &  1.08  &  0.04     \\
\hline
11930  &  2001-10-07  &  1.11  &  0.05  &  1.00  &  0.04     \\
\hline
14652  &  2001-10-05  &  0.62  &  0.05  &  0.61  &  0.05     \\
\hline
15656  &  2001-10-03  &  0.77  &  0.04  &  0.62  &  0.04     \\
\hline
15694  &  2001-10-08  &  1.06  &  0.05  &  1.01  &  0.04     \\
\hline
16058  &  1998-10-25  &  0.55  &  0.03  &  0.42  &  0.02     \\
\hline
17506  &  1998-11-29  &  0.64  &  0.06  &  0.61  &  0.05     \\
\hline
19349  &  2001-10-05  &  1.04  &  0.04  &  0.98  &  0.03     \\
\hline
19476  &  1998-10-26  &  1.16  &  0.04  &  1.22  &  0.04     \\
\hline
20610  &  1998-11-28  &  1.19  &  0.08  &  1.15  &  0.07     \\
\hline
20893  &  2001-10-08  &  0.91  &  0.04  &  1.11  &  0.04     \\
\hline
21051  &  1998-11-26  &  1.13  &  0.07  &  1.18  &  0.06     \\
\hline
21552  &  2001-10-03  &  1.01  &  0.03  &  0.94  &  0.03     \\
\hline
26076  &  1998-11-27  &  1.30  &  0.10  &  1.28  &  0.09     \\
\hline
26846  &  2001-10-03  &  1.11  &  0.04  &  1.10  &  0.03     \\
\hline
27278  &  1998-11-26  &  1.03  &  0.05  &  1.08  &  0.05     \\
\hline
27371  &  1998-11-28  &  1.09  &  0.03  &  1.10  &  0.03     \\
\hline
27697  &  1998-11-29  &  1.14  &  0.08  &  1.24  &  0.07     \\
       &  1998-11-30  &  1.13  &  0.04  &  1.19  &  0.04     \\
\hline
28305  &  1998-11-30  &  1.18  &  0.04  &  1.12  &  0.03     \\
       &  2006-10-03  &  1.20  &  0.04  & \nodata & \nodata    \\
       &  2007-02-01  &  1.14  &  0.04  &  1.02  &  0.03     \\
\hline
28307  &  1998-11-30  &  1.07  &  0.03  &  1.13  &  0.02     \\
\hline
32357  &  1998-11-26  &  1.05  &  0.03  &  1.12  &  0.02     \\
\hline
32887  &  2001-10-08  &  0.94  &  0.03  &  0.96  &  0.03     \\
\hline
33856  &  1998-11-30  &  1.07  &  0.06  &  1.17  &  0.05     \\
\hline
35186  &  2001-10-07  &  1.22  &  0.04  &  1.23  &  0.04     \\
\hline
42633  &  2001-10-08  &  1.04  &  0.05  &  1.07  &  0.05     \\
\hline
42995  &  1998-11-27  &  0.38  &  0.05  &  0.39  &  0.04     \\
\hline
49161  &  1998-11-28  &  1.08  &  0.06  &  0.99  &  0.04     \\
       &  2001-10-07  &  1.02  &  0.03  &  0.93  &  0.03     \\
       &  2006-10-04  &  0.93  &  0.13  &  1.03  &  0.03     \\
       &  2007-02-01  &  0.97  &  0.13  &  0.90  &  0.01     \\
\hline
50522  &  1998-11-26  &  1.11  &  0.05  &  1.16  &  0.04     \\
\hline
57478  &  1998-11-27  &  1.14  &  0.07  &  1.14  &  0.06     \\
\hline
59148  &  1998-11-28  &  1.09  &  0.06  &  1.00  &  0.05     \\
\hline
62345  &  1998-11-30  &  1.21  &  0.03  &  1.19  &  0.03     \\
       &  2006-10-03  &  1.09  &  0.04  &  1.18  &  0.03     \\
\hline
62898  &  1998-11-26  &  0.58  &  0.06  &  0.62  &  0.05     \\
\hline
74874  &  1998-11-28  &  1.03  &  0.05  &  1.05  &  0.04     \\
\hline
78235  &  1998-11-28  &  1.11  &  0.04  &  1.11  &  0.03     \\
       &  2007-03-30  &  1.08  &  0.04  &  1.15  &  0.04     \\
\hline
82635  &  1998-11-30  &  1.05  &  0.02  &  1.08  &  0.02     \\
       &  2007-03-12  &  1.07  &  0.07  &  1.02  &  0.07     \\
       &  2007-03-30  &  1.03  &  0.03  &  1.04  &  0.03     \\
\hline
85444  &  1998-11-30  &  1.11  &  0.05  &  1.23  &  0.06     \\
       &  2007-03-30  &  1.09  &  0.04  &  0.99  &  0.07     \\
\hline
93813  &  1998-11-30  &  1.16  &  0.06  &  1.17  &  0.05     \\
       &  2007-03-11  &  1.08  &  0.06  &  1.17  &  0.14     \\
\hline
98262  &  2005-06-27  &  1.05  &  0.06  &  1.01  &  0.06     \\
       &  2007-05-27  &  1.01  &  0.03  &  1.00  &  0.04     \\
\hline
105707  &  2004-05-28  &  1.14  &  0.04  &  1.01  &  0.04     \\
        &  2004-06-29  &  0.99  &  0.06  &  1.14  &  0.11     \\
        &  2005-06-26  &  1.15  &  0.05  &  0.93  &  0.04     \\
        &  2008-06-22  &  1.11  &  0.02  &  1.04  &  0.02     \\
\hline
124294  &  2004-05-28  &  1.12  &  0.04  &  0.92  &  0.04     \\
        &  2005-06-25  &  0.98  &  0.05  &  0.93  &  0.06     \\
        &  2007-05-28  &  1.08  &  0.04  &  0.96  &  0.08     \\
        &  2008-06-22  &  0.98  &  0.03  &  0.90  &  0.04     \\
\hline
124547  &  2004-05-28  &  1.19  &  0.05  &  0.85  &  0.06     \\
        &  2004-06-29  &  1.00  &  0.05  &  1.00  &  0.06     \\
        &  2005-06-26  &  0.98  &  0.05  &  0.90  &  0.06     \\
\hline
131430  &  2004-07-01  &  1.28  &  0.15  & \nodata & \nodata    \\
\hline
139063  &  2004-05-28  &  0.92  &  0.05  &  0.81  &  0.06     \\
        &  2004-06-30  &  1.05  &  0.09  &  1.05  &  0.12     \\
        &  2004-07-21  &  1.20  &  0.08  &  1.07  &  0.12     \\
\hline
140227  &  2004-07-20  &  1.04  &  0.07  &  0.97  &  0.03     \\
        &  2007-05-28  &  1.24  &  0.07  &  1.08  &  0.18     \\
\hline
145849  &  2004-06-29  &  0.63  &  0.05  & \nodata & \nodata     \\
\hline
153727  &  2004-07-21  &  1.11  &  0.18  & \nodata & \nodata   \\
\hline
153834  &  2004-07-01  &  1.32  &  0.05  & \nodata & \nodata   \\
\hline
156283  &  2004-09-03  &  1.14  &  0.03  &  1.08  &  0.05     \\
        &  2005-06-27  &  1.19  &  0.05  &  0.89  &  0.03     \\
        &  2007-03-11  &  0.95  &  0.04  &  1.02  &  0.05     \\
        &  2007-05-27  &  1.11  &  0.04  &  1.00  &  0.07     \\
        &  2008-06-22  &  1.07  &  0.01  &  0.99  &  0.02     \\
\hline
161832  &  2004-06-30  &  1.12  &  0.04  & \nodata & \nodata    \\
\hline
163770  &  2004-05-27  &  0.97  &  0.03  &  0.82  &  0.04     \\
        &  2005-06-26  &  0.93  &  0.03  &  0.72  &  0.05     \\
\hline
168454  &  2004-05-27  &  1.18  &  0.11  &  1.25  &  0.12     \\
        &  2004-07-20  &  1.16  &  0.06  &  1.13  &  0.08     \\ 
        &  2004-09-03  &  1.32  &  0.08  &  1.11  &  0.08     \\
        &  2005-06-25  &  1.15  &  0.07  &  1.08  &  0.08     \\
        &  2007-05-27  &  1.12  &  0.05  &  0.94  &  0.06     \\
        &  2008-06-22  &  1.17  &  0.06  &  1.10  &  0.08     \\
\hline
171443  &  2004-05-28  &  0.99  &  0.05  &  0.94  &  0.04     \\
        &  2004-06-29  &  1.01  &  0.06  &  0.94  &  0.05     \\
        &  2005-06-25  &  1.12  &  0.04  &  1.00  &  0.05     \\
        &  2006-08-31  &  1.01  &  0.05  &  1.01  &  0.04     \\
\hline
175443  &  2004-07-20  &  1.12  &  0.06  &  1.37  &  0.08     \\
        &  2005-06-26  &  1.10  &  0.18  &  0.97  &  0.07     \\
\hline
177199  &  2004-07-21  &  0.82  &  0.16  & \nodata & \nodata    \\
\hline
183387  &  2004-07-01  &  1.30  &  0.08  & \nodata & \nodata    \\
\hline
183439  &  2001-10-07  &  0.67  &  0.07  &  0.65  &  0.06     \\
\hline
184835  &  2001-10-03  &  1.23  &  0.07  &  1.14  &  0.06     \\
\hline
187372  &  1998-11-29  &  0.68  &  0.10  &  0.56  &  0.08     \\
\hline
190327  &  1998-11-28  &  1.10  &  0.06  &  1.11  &  0.05     \\
\hline
190940  &  2001-10-08  &  1.02  &  0.04  &  1.04  &  0.04     \\
        &  2004-05-28  &  1.25  &  0.05  &  1.04  &  0.04     \\
        &  2004-06-29  &  1.16  &  0.04  &  0.97  &  0.03     \\
        &  2005-06-25  &  1.34  &  0.07  &  1.15  &  0.07     \\
        &  2006-08-31  &  1.02  &  0.03  &  1.03  &  0.04     \\
        &  2008-06-22  &  1.09  &  0.03  &  1.04  &  0.04     \\
\hline
194317  &  2005-06-27  &  1.26  &  0.06  & \nodata & \nodata    \\
\hline
196574  &  1998-11-26  &  1.17  &  0.06  &  1.17  &  0.05     \\
\hline
198026  &  2001-10-04  &  0.91  &  0.04  &  0.76  &  0.04     \\
\hline
198134  &  2001-10-08  &  1.21  &  0.05  &  1.13  &  0.04     \\
        &  2004-06-30  &  1.22  &  0.06  &  1.05  &  0.06     \\
        &  2006-08-31  &  1.10   &  0.05  &  1.06  &  0.05     \\
\hline
199169  &  2001-10-03  &  1.06  &  0.03  &  0.85  &  0.03     \\
\hline
199253  &  1998-11-27  &  1.06  &  0.04  &  1.05  &  0.04     \\
\hline
200644  &  2001-10-04  &  0.75  &  0.05  &  0.63  &  0.05     \\
\hline
202951  &  1998-10-26  &  0.90  &  0.06  &  0.95  &  0.05     \\
\hline
203387  &  2006-10-05  &  0.92  &  0.03  &  1.01  &  0.04     \\
        &  2008-06-22  &  1.06  &  0.02  &  1.06  &  0.02     \\
\hline
204724  &  2001-10-05  &  0.70  &  0.02  &  0.58  &  0.02     \\
\hline
205435  &  1998-10-27  &  1.17  &  0.02  &  1.12  &  0.02     \\
        &  2006-10-05  &  1.05  &  0.03  &  1.07  &  0.03     \\
\hline
209747  &  2001-10-07  &  0.97  &  0.06  &  0.76  &  0.05     \\
\hline
210434  &  1998-10-25  &  1.15  &  0.05  &  1.13  &  0.04     \\
\hline
211073  &  2001-10-03  &  1.06  &  0.02  &  0.98  &  0.02     \\
        &  2004-06-29  &  1.04  &  0.03  &  0.91  &  0.03     \\
        &  2004-07-20  &  1.04  &  0.07  &  1.02  &  0.05     \\
        &  2005-06-26  &  1.20  &  0.04  &  0.99  &  0.03     \\
        &  2006-08-31  &  1.06  &  0.03  &  0.97  &  0.04     \\
        &  2008-06-23  &  1.06  &  0.03  &  0.97  &  0.04     \\
\hline
211391  &  2006-10-03  &  1.16  &  0.06  &  1.18  &  0.03     \\
        &  2008-06-23  &  1.16  &  0.02  &  1.15  &  0.03     \\
\hline
214868  &  2004-06-30  &  1.22  &  0.05  &  1.02  &  0.05     \\
        &  2005-06-26  &  1.31  &  0.06  &  1.10  &  0.04     \\
        &  2006-09-01  &  1.12  &  0.04  &  1.03  &  0.05     \\
\hline
215167  &  2001-10-07  &  1.00  &  0.04  &  0.82  &  0.03     \\
        &  2004-07-01  &  1.10  &  0.05  &  0.92  &  0.05     \\
        &  2005-06-27  &  0.88  &  0.16  &  0.72  &  0.08     \\
        &  2006-09-01  &  0.97  &  0.05  &  0.89  &  0.06     \\
\hline
216446  &  2001-10-08  &  1.14  &  0.04  &  1.14  &  0.04     \\
\hline
216718  &  1998-11-26  &  1.08  &  0.08  &  1.13  &  0.07     \\
\hline
217382  &  2001-10-08  &  1.06  &  0.04  &  1.04  &  0.04     \\
\hline
217906  &  2001-10-03  &  0.81  &  0.01  &  0.73  &  0.01     \\
\hline
218452  &  2001-10-07  &  1.10  &  0.03  &  1.10  &  0.03     \\
\hline
218527  &  1998-11-26  &  1.24  &  0.06  &  1.30  &  0.06     \\
\hline
219916  &  1998-10-26  &  1.05  &  0.06  &  1.19  &  0.05     \\
\hline
220363  &  2001-10-07  &  1.10  &  0.04  &  1.10  &  0.04     \\
\hline
221673  &  2004-07-20  &  1.09  &  0.11  &  0.91  &  0.04     \\
\hline
222404  &  1998-10-26  &  1.10  &  0.04  &  1.08  &  0.03     \\
        &  2006-09-01  &  1.10  &  0.04  &  1.09  &  0.04     \\
        &  2006-10-04  &  1.28  &  0.05  &  1.25  &  0.04     \\
\hline
222643  &  2001-10-03  &  1.16  &  0.09  & \nodata & \nodata    \\
\hline
222764  &  2001-10-04  &  0.82  &  0.04  &  0.71  &  0.04     \\
\hline
223460  &  1998-11-29  &  1.07  &  0.03  &  1.07  &  0.03     \\
\hline
224533  &  1998-10-26  &  1.07  &  0.03  &  1.13  &  0.03     \\ 
\enddata
\end{deluxetable}

\begin{deluxetable}{lrrrrcrc}
\tablecaption{ Summary of Sample }
\tablehead{
\colhead{Star} &
\colhead{$B-V$ } &
\colhead{$M_V$} &
\colhead{$\langle V/R \rangle (\sigma)$} &
\colhead{$\langle V/R \rangle (\sigma)$} &
\colhead{[Fe/H]\tablenotemark{a}} &
\colhead{$v_{r}$} & 
\colhead{$A_V$} \\
\colhead{(HD)} &
\colhead{} &
\colhead{(mag)} &
\colhead{(H)} &
\colhead{(K)} &
\colhead{} &
\colhead{km s$^{-1}$} &
\colhead{mag} }
\startdata
1013   &  1.573   &  -0.2  & 0.79  (0.03)  & 0.70  (0.03)  &       \nodata    & -46.6   &  0.029 \\
1227   &  0.917  &  0.45  & 1.15  (0.07)  & 1.13  (0.06)  &       \nodata    &    1.4   &  0.052 \\
1522   &  1.213  &  -1.2  & 1.14  (0.05)  & 1.03  (0.03)  &      --0.09      &   18.6   &  0.020 \\
3627   &  1.278  &  0.8  & 1.096 (0.018) & 1.094 (0.024) &        0.04      &  -7.3   &  0.000 \\
4128   &  1.023  &  -0.3  & 1.049 (0.017) & 1.095 (0.021) &      --0.09      &   13.0   &  0.000 \\
4482   &  0.973  &  0.98  & 1.11  (0.03)  & 1.10  (0.03)  &       \nodata    &  -0.8   &  0.004 \\
4730   &  1.312  &  0.59  & 1.06  (0.06)  & 1.06  (0.05)  &      --0.03      &    3.5   &  0.034 \\
5516   &  0.940  &  0.06  & 1.10  (0.04)  & 1.22  (0.03)  &      --0.54      & -10.3   &  0.000 \\
6953   &  1.466  &  0.59  & 1.14  (0.05)  & 1.08  (0.04)  &       \nodata    &    3.2   &  0.017 \\
7318   &  1.035  &  -0.67  & 1.16  (0.06)  & 1.14  (0.05)  &      --0.08      &    5.9   &  0.020 \\
9138   &  1.372  &  -0.37  & 0.84  (0.041) & 0.928 (0.037) &      --0.39      &   34.2   &  0.033 \\
9774   &  0.960  &  -0.4  & 1.16  (0.06)  & 1.18  (0.06)  &      --0.1       &  -4.0   &  0.062 \\
11559   &  0.941  &  0.79  & 1.13  (0.05)  & 1.08  (0.04)  &      --0.11      &   30.3   &  0.000 \\
11930   &  1.430  &  -0.55  & 1.11  (0.05)  & 1.00  (0.04)  &      --0.18      &   26.7   &  0.031 \\
14652   &  1.652  &  -2.73  & 0.62  (0.05)  & 0.61  (0.05)  &       \nodata    &   22.8   &  0.094 \\
15656   &  1.470  &  -0.24  & 0.77  (0.04)  & 0.62  (0.04)  &      --0.16      & -35.9   &  0.057 \\
15694   &  1.255  &  -0.56  & 1.06  (0.05)  & 1.01  (0.04)  &      --0.17      &   25.6   &  0.021 \\
16058   &  1.652  &  \nodata  & 0.55  (0.03)  & 0.42  (0.02)  &       \nodata    &  -7.1   &  \nodata \\
17506   &  1.690  &  -4.29  & 0.64  (0.06)  & 0.61  (0.05)  &       \nodata    &  -1.1   &  0.573 \\
19349   &  1.595  &  -0.81  & 1.04  (0.04)  & 0.98  (0.03)  &       \nodata    &   16.7   &  0.041 \\
19476   &  0.980  &  1.12  & 1.16  (0.04)  & 1.22  (0.04)  &        0.04      &   29.2   &  0.000 \\
20610   &  0.904  &  0.41  & 1.19  (0.08)  & 1.15  (0.07)  &      --0.07      &   23.9   &  0.003 \\
20893   &  1.229  &  0.14  & 0.91  (0.04)  & 1.11  (0.04)  &        0.02      &    3.0   &  0.020 \\ 
21051   &  1.230  &  0.28  & 1.13  (0.07)  & 1.18  (0.06)  &       \nodata    &   18.6   &  0.004 \\
21552   &  1.367  &  -0.81  & 1.01  (0.03)  & 0.94  (0.03)  &      --0.24      &   14.4   &  0.037 \\
26076   &  1.012  &  0.78  & 1.30  (0.10)  & 1.28  (0.09)  &       \nodata    &  -5.5   &  0.023 \\
26846   &  1.162  &  0.86  & 1.11  (0.04)  & 1.10  (0.03)  &        0.09      &    6.8   &  0.000 \\
27278   &  0.955  &  0.79  & 1.03  (0.05)  & 1.08  (0.05)  &       \nodata    &   24.1   &  0.049 \\
27371   &  0.987  &  0.28  & 1.09  (0.03)  & 1.10  (0.03)  &      --0.02      &   38.7   &  0.000 \\
27697   &  0.983  &  0.4  & 1.132 (0.036) & 1.202 (0.035) &        0         &   38.8   &  0.000 \\
28305   &  1.014  &  0.15  & 1.173 (0.023) & 1.070 (0.021) &        0.04      &   38.2   &  0.000 \\
28307   &  0.951  &  0.42  & 1.07  (0.03)  & 1.13  (0.02)  &        0.04      &   39.8   &  0.000 \\
32357   &  1.132  &  -0.33  & 1.05  (0.03)  & 1.12  (0.02)  &       \nodata    &  -8.0   &  0.210 \\
32887   &  1.472  &  -1.04  & 0.94  (0.03)  & 0.96  (0.03)  &      --0.17      &    1.0   &  0.000 \\
33856   &  1.185  &  -0.65  & 1.07  (0.06)  & 1.17  (0.05)  &        0.06      &   41.0   &  0.025 \\
35186   &  1.416  &  -0.99  & 1.22  (0.04)  & 1.23  (0.04)  &      --0.03      & -19.7   &  0.095 \\
42633   &  1.340  &  -0.57  & 1.04  (0.05)  & 1.07  (0.05)  &       0          &    8.6   &  0.089 \\  
42995   &  1.586  &  -1.87  & 0.38  (0.05)  & 0.39  (0.04)  &       \nodata    &   19.0   &  0.032 \\
49161   &  1.393  &  -1.1  & 1.026 (0.026) & 0.918 (0.009) &      --0.03      &   46.2   &  0.049 \\ 
50522   &  0.849  &  0.76  & 1.11  (0.05)  & 1.16  (0.04)  &       0.05       &    8.9   &  0.000 \\
57478   &  0.974  &  -0.71  & 1.14  (0.07)  & 1.14  (0.06)  &       \nodata    &   13.2   &  0.039 \\
59148   &  1.117  &  -0.34  & 1.09  (0.06)  & 1.00  (0.05)  &       \nodata    &   36.0   &  0.000 \\
62345   &  0.932  &  0.35  & 1.167 (0.024) & 1.185 (0.021) &      --0.16      &   20.6   &  0.000 \\  
62898   &  1.594  &  -1.04  & 0.58  (0.06)  & 0.62  (0.05)  &       \nodata    & -13.4   &  0.033 \\
74874   &  0.684  &  0.29  & 1.03  (0.05)  & 1.05  (0.04)  &      --0.14      &   36.4   &  0.000 \\
78235   &  0.888  &  0.95  & 1.095 (0.028) & 1.124 (0.024) &       \nodata    & -13.1   &  0.000 \\
82635   &  0.916  &  0.89  & 1.045 (0.016) & 1.065 (0.016) &      --0.15      & -11.7   &  0.000 \\
85444   &  0.928  &  -0.51  & 1.098 (0.031) & 1.128 (0.046) &      --0.14      & -14.5   &  0.006 \\
93813   &  1.245  &  -0.14  & 1.120 (0.042) & 1.170 (0.074) &      --0.3       &  -1.2   &  0.000 \\
98262   &  1.398  &  -2.08  & 1.018 (0.027) & 1.003 (0.033) &      --0.2       &  -9.6   &  0.015 \\
105707   &  1.331  &  -1.84  & 1.110 (0.016) & 1.019 (0.016) &      --0.13      &    4.9   &  0.016 \\
124294   &  1.336  &  0  & 1.037 (0.019) & 0.918 (0.024) &      --0.39      &  -4.0   &  0.000 \\ 
124547   &  1.368  &  -1.06  & 1.057 (0.029) & 0.917 (0.035) &       0.17       &   10.5   &  0.048 \\
131430   &  1.331  &  \nodata  & 1.28  (0.15)  & \nodata       &       0.04       &    8.8   &  \nodata \\
139063   &  1.380  &  -0.28  & 1.008 (0.038) & 0.893 (0.049) &      --0.18      & -24.9   &  0.000 \\
140227   &  1.362  &  -1.55  & 1.140 (0.049) & 0.973 (0.030) &       \nodata    & -30.4   &  0.060 \\
145849   &  1.340  &  \nodata  & 0.63  (0.05)  & \nodata       &       \nodata    & -30.6   &  \nodata \\
153727   &  1.355  &  \nodata  & 1.11  (0.18)  & \nodata       &       \nodata    &   44.2   &  \nodata \\
153834   &  \nodata  &  \nodata  & 1.32  (0.05)  & \nodata       &       \nodata    &   11.3   &  \nodata \\  
156283   &  1.437  &  -2.09  & 1.076 (0.009) & 0.977 (0.015) &      --0.18      & -25.6   &  0.041 \\
161832   &  1.390  &  \nodata  & 1.12  (0.04)  &  \nodata  &       \nodata    & -32     &  \nodata \\
163770   &  1.351  &  -2.7  & 0.950 (0.021) & 0.781 (0.031) &      --0.24      & -28.3   &  0.080 \\
168454   &  1.376  &  -2.16  & 1.168 (0.027) & 1.068 (0.032) &      --0.01      & -19.9   &  0.024 \\
171443   &  1.333  &  0.21  & 1.046 (0.024) & 0.973 (0.022) &      --0.18      &   35.8   &  0.000 \\
175443   &  1.349  &  0.15  & 1.118 (0.057) & 1.143 (0.053) &       \nodata    &   13.3   &  0.036 \\
177199   &  1.339  &  \nodata  & 0.82  (0.16)  & \nodata       &       \nodata    &  -7.0   &  \nodata \\
183387   &  1.318  &  \nodata  & 1.30  (0.08)  & \nodata       &       \nodata    &  \nodata &  \nodata \\
183439   &  1.502  &  -0.35  & 0.67  (0.07)  & 0.65  (0.06)  &       \nodata    & -85.5   &  0.014 \\
184835   &  1.242  &  0  & 1.23  (0.07)  & 1.14  (0.06)  &       \nodata    &  -7.3   &  0.212 \\
187372   &  1.640  &  -1.76  & 0.68  (0.10)  & 0.56  (0.08)  &       \nodata    &    3     &  0.370 \\
190327   &  1.062  &  -0.52  & 1.10  (0.06)  & 1.11  (0.05)  &      --0.15      & -29.5   &  0.177 \\
190940   &  1.313  &  -0.97  & 1.099 (0.016) & 1.024 (0.016) &       0.03       &  -9.8   &  0.027 \\
194317   &  1.330  &  \nodata  & 1.26  (0.06)  & \nodata       &      --0.17      & -14.6   &  \nodata \\
196574   &  0.949  &  -1.03  & 1.17  (0.06)  & 1.17  (0.05)  &      --0.13      &  -5.6   &  0.065 \\ 
198026   &  1.645  &  -1.24  & 0.91  (0.04)  & 0.76  (0.04)  &       \nodata    & -22.0   &  0.090 \\
198134   &  1.299  &  -0.5  & 1.172 (0.030) & 1.091 (0.028) &      --0.12      & -24.1   &  0.038 \\
199169   &  1.480  &  -1.76  & 1.06  (0.03)  & 0.85  (0.03)  &      --0.16      &    8.1   &  0.115 \\
199253   &  1.119  &  -0.64  & 1.06  (0.04)  & 1.05  (0.04)  &      --0.18      & -11.2   &  0.049 \\
200644   &  1.650  &  -1.02  & 0.75  (0.05)  & 0.63  (0.05)  &       \nodata    & -15.3   &  0.043 \\
202951   &  1.648  &  -1.97  & 0.90  (0.06)  & 0.95  (0.05)  &       \nodata    & -37.0   &  0.099 \\
203387   &  0.899  &  0.18  & 1.017 (0.017) & 1.050 (0.018) &      --0.23      &   11.5   &  0.000 \\
204724   &  1.618  &  -1.11  & 0.70  (0.02)  & 0.58  (0.02)  &       \nodata    & -18.9   &  0.050 \\
205435   &  0.887  &  1.1  & 1.13  (0.017) & 1.105 (0.017) &      --0.31      &    6.9   &  0.000 \\
209747   &  1.452  &  0.3  & 0.97  (0.06)  & 0.76  (0.05)  &       0.02       & -18.9   &  0.005 \\
210434   &  0.983  &  1.33  & 1.15  (0.05)  & 1.13  (0.04)  &       \nodata    & -18.1   &  0.006 \\
211073   &  1.385  &  -1.69  & 1.069 (0.012) & 0.970 (0.013) &       0.02       & -10.6   &  0.094 \\
211391   &  0.983  &  0.32  & 1.160 (0.030) & 1.165 (0.021) &       0.01       & -14.7   &  0.000 \\
214868   &  1.320  &  -0.35  & 1.191 (0.028) & 1.058 (0.026) &      --0.25      & -11.0   &  0.019 \\
215167   &  1.371  &  -1.18  & 1.016 (0.026) & 0.843 (0.023) &      --0.23      &   21.6   &  0.078 \\
216446   &  1.257  &  -0.23  & 1.14  (0.04)  & 1.14  (0.04)  &      --0.19      & -31.9   &  0.030 \\
216718   &  0.880  &  0.95  & 1.08  (0.08)  & 1.13  (0.07)  &       0.04       &  -8.8   &  0.015 \\
217382   &  1.417  &  -0.67  & 1.06  (0.04)  & 1.04  (0.04)  &      --0.11      &    2.9   &  0.049 \\
217906   &  1.668  &  -1.47  & 0.81  (0.01)  & 0.73  (0.01)  &       \nodata    &    8.0   &  0.000 \\
218452   &  1.409  &  0.22  & 1.10  (0.03)  & 1.10  (0.03)  &      --0.02      & -11.6   &  0.032 \\
218527   &  0.906  &  0.73  & 1.24  (0.06)  & 1.30  (0.06)  &      --0.31      & -17.8   &  0.009 \\
219916   &  0.837  &  0.7  & 1.05  (0.06)  & 1.19  (0.05)  &      --0.07      & -18.4   &  0.000 \\
220363   &  1.316  &  0.07  & 1.10  (0.04)  & 1.10  (0.04)  &      --0.01      &  -3.9   &  0.018 \\
221673   &  1.385  &  -1.16  & 1.09  (0.11)  & 0.91  (0.04)  &      --0.03      & -24.7   &  0.090 \\
222404   &  1.030  &  2.51  & 1.144 (0.025) & 1.128 (0.021) &       0          & -42.4   &  0.000 \\
222643   &  1.362  &  \nodata  & 1.16  (0.09)  & \nodata       &       0.02       &    6.8   &  \nodata \\
222764   &  1.686  &  -1.95  & 0.82  (0.04)  & 0.71  (0.04)  &       \nodata    & -32.7   &  0.060 \\
223460   &  0.793  &  0.25  & 1.07  (0.03)  & 1.07  (0.03)  &       \nodata    &    0.7   &  0.086 \\
224533   &  0.930  &  0.68  & 1.07  (0.03)  & 1.13  (0.03)  &      --0.13      &  -0.2   &  0.000 \\
\enddata
\tablenotetext{a}{Metallicities taken from McWilliam (1990).}
\end{deluxetable}

\clearpage
\epsscale{0.8}
\plotone{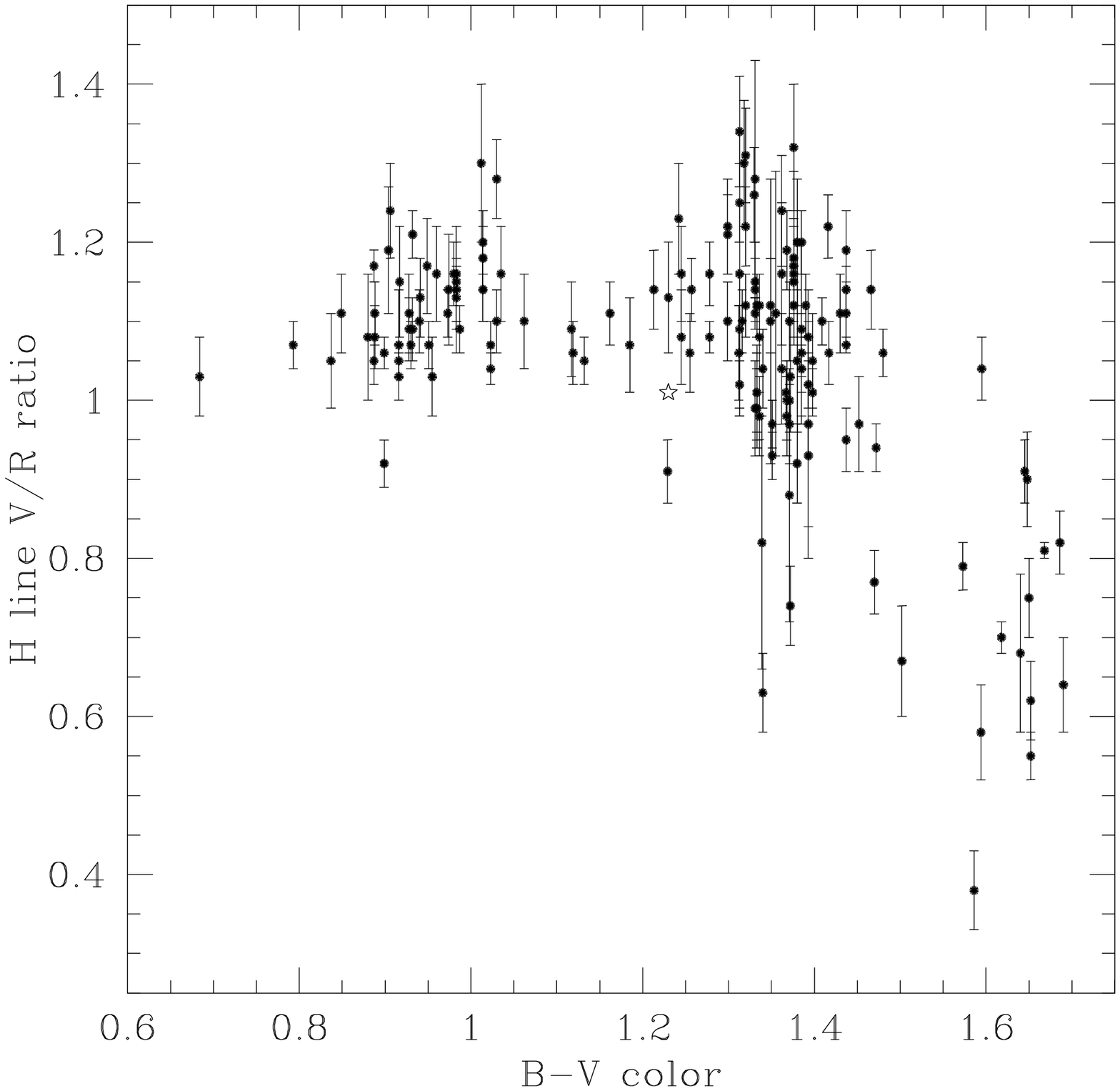}
\figcaption{The $V/R$ value of the \ion{Ca}{2} H line from Table 1 is plotted 
versus the $B-V$ color (Table 2). The single star symbol represents Arcturus.
Where multiple observations are available for a star, the result for each 
individual spectrum is shown as a separate data point. Thus a number of stars
are represented by multiple points. 
\label{fig1}}

\clearpage
\epsscale{0.8}
\plotone{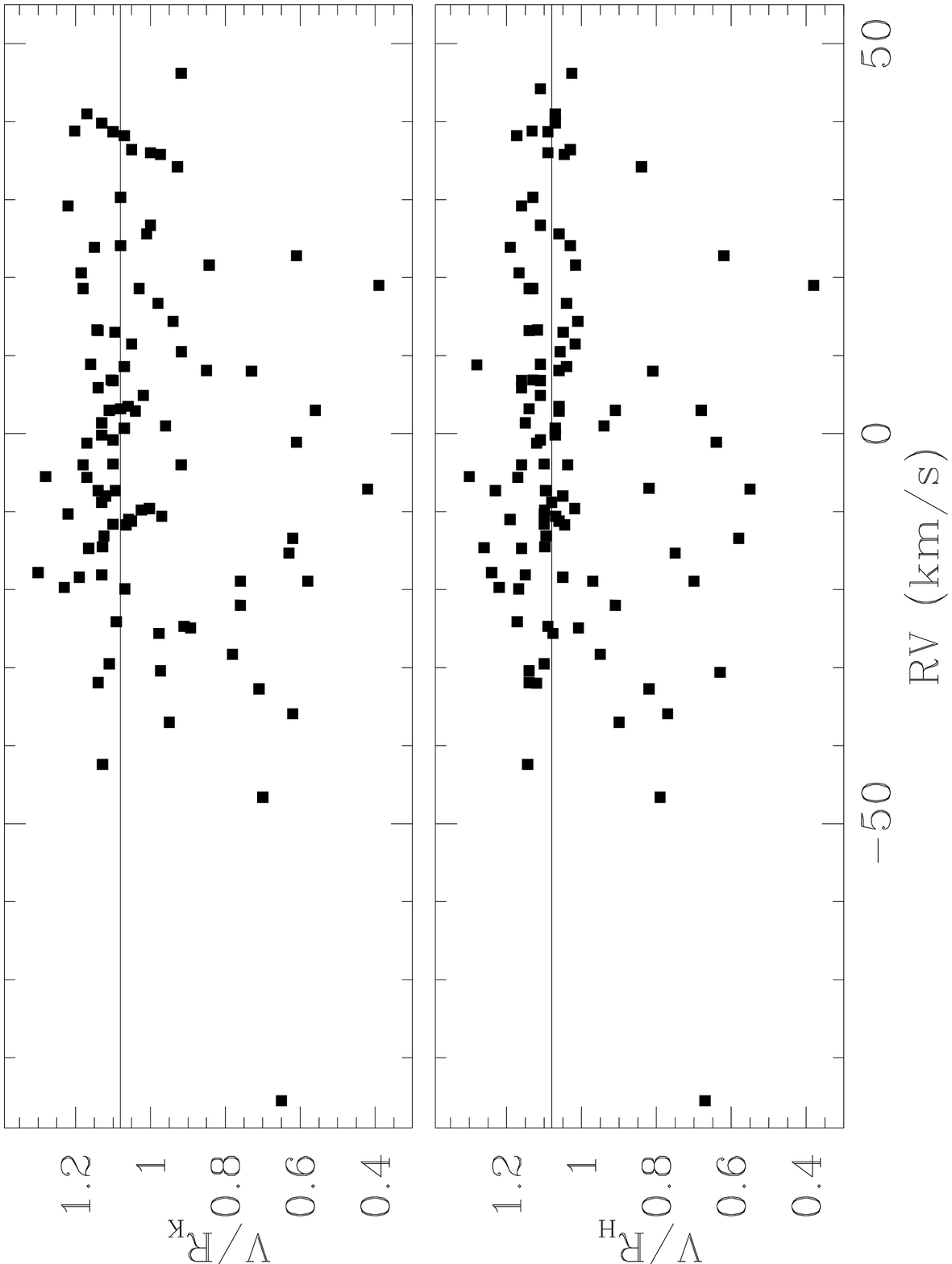}
\figcaption{The H and K line asymmetry parameters $(V/R)_K$ and $(V/R)_H$ 
versus heliocentric radial velocity. The lack of correlations in this figure 
is presented as evidence that the asymmetry trends seen in this, and previous
papers, are not dominated by the effects of interstellar absorption. 
\label{fig2}}

\clearpage
\epsscale{0.8}
\plotone{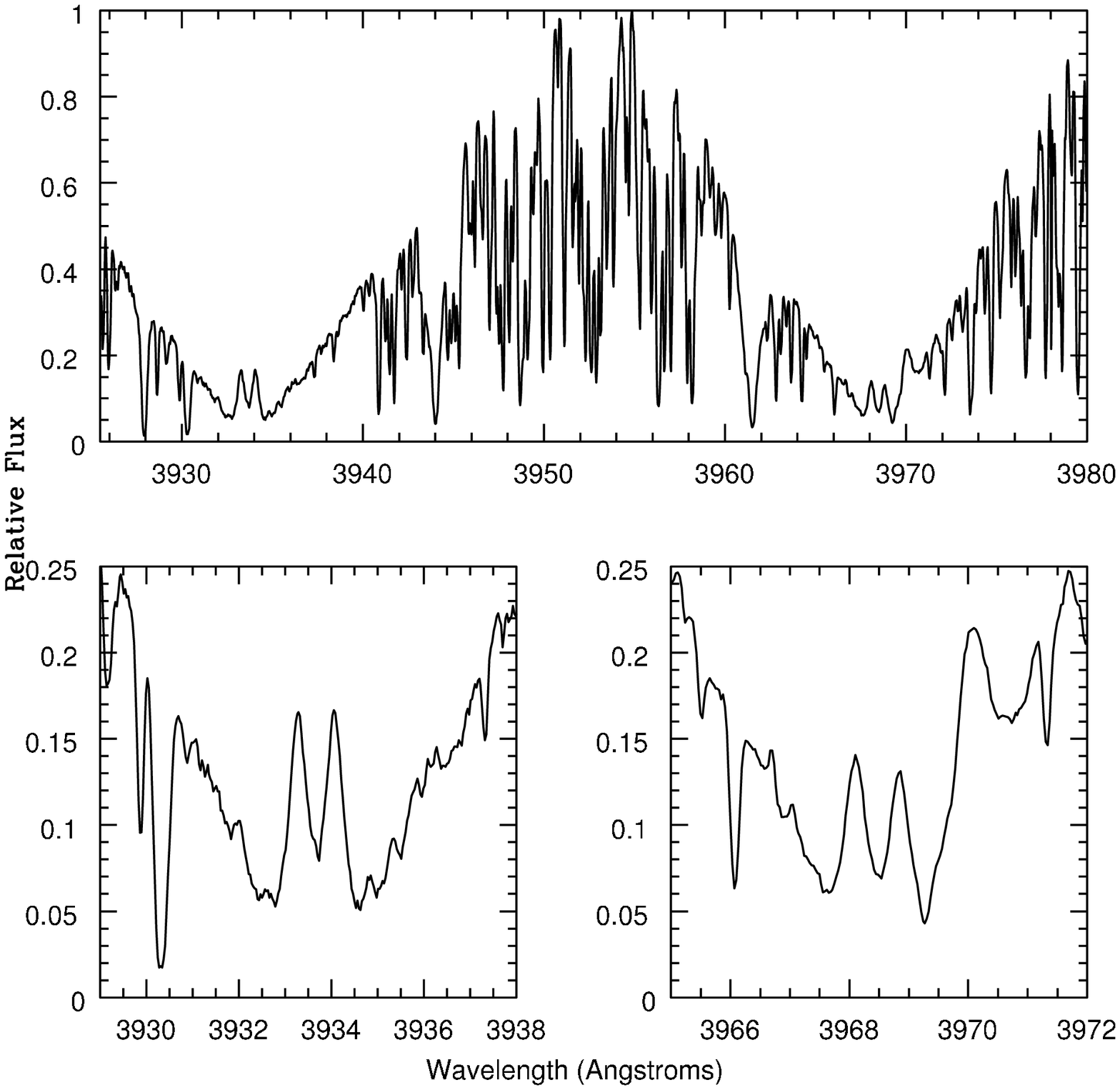}
\figcaption{The order of the normalized spectrum of HD 156283 that contains the
\ion{Ca}{2} H and K lines is plotted in the upper panel. This spectrum was 
observed on 22 June 2008 using the 2.7 m telescope at McDonald Observatory. In
the bottom left window, an expanded version of the core of the K line is 
plotted, to clearly show the near symmetry of the emission peaks. The core of 
the H line is plotted in the bottom right window, and reveals the blueward
emission peak to be stronger than the redward peak.
\label{fig3}}

\clearpage
\epsscale{0.8}
\plotone{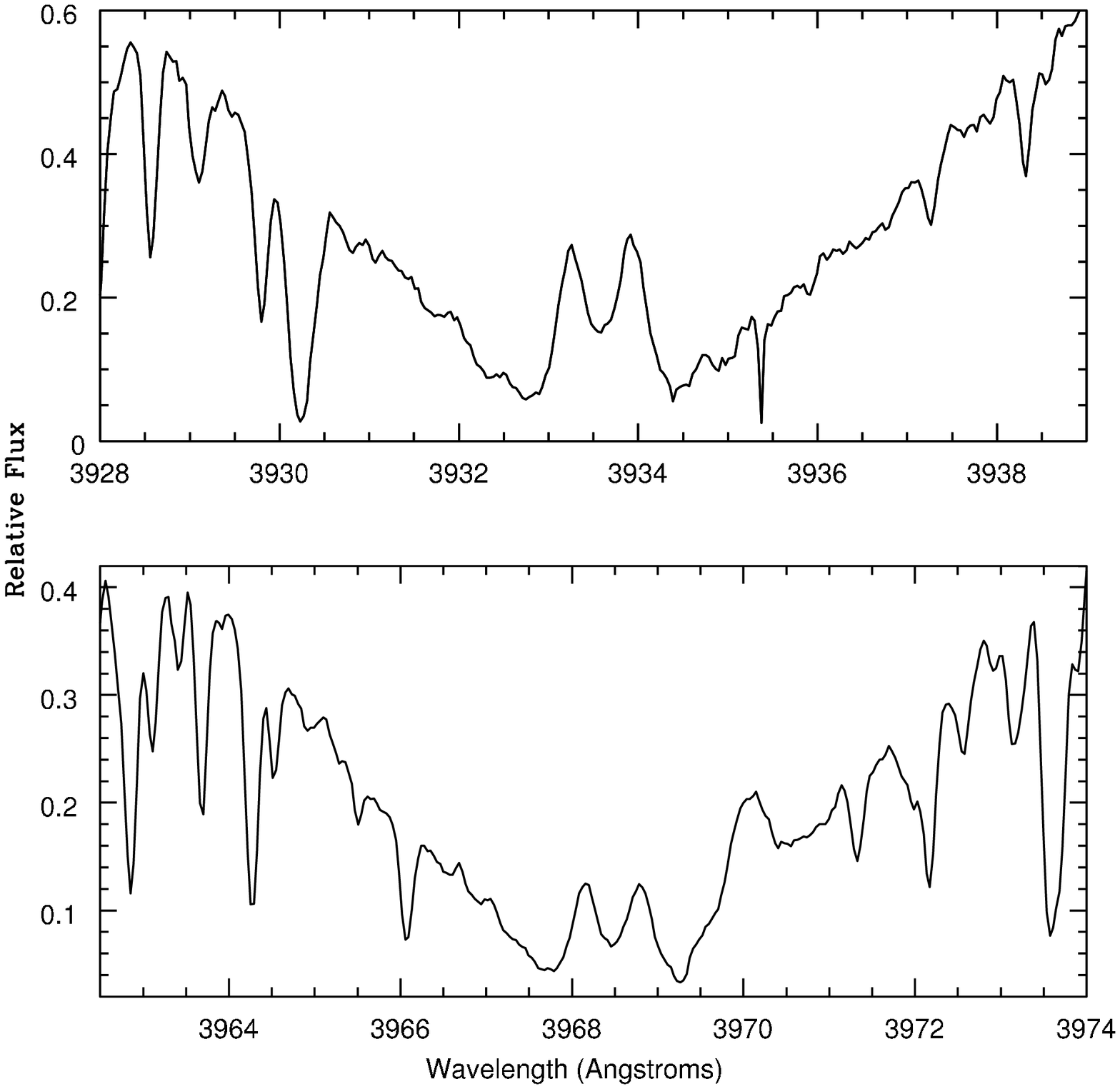}
\figcaption{The echelle order of the normalized spectrum of HD 21552 containing
the \ion{Ca}{2} K line is plotted in the top panel, with the H-line order 
shown in the bottom window. This spectrum was observed on 3 October 2001 using 
the 2.1 m telescope at McDonald Observatory. Whereas the H emission line is
nearly symmetric, the K line shows a slightly red-enhanced emission profile.
\label{fig4}}

\clearpage
\epsscale{0.8}
\plotone{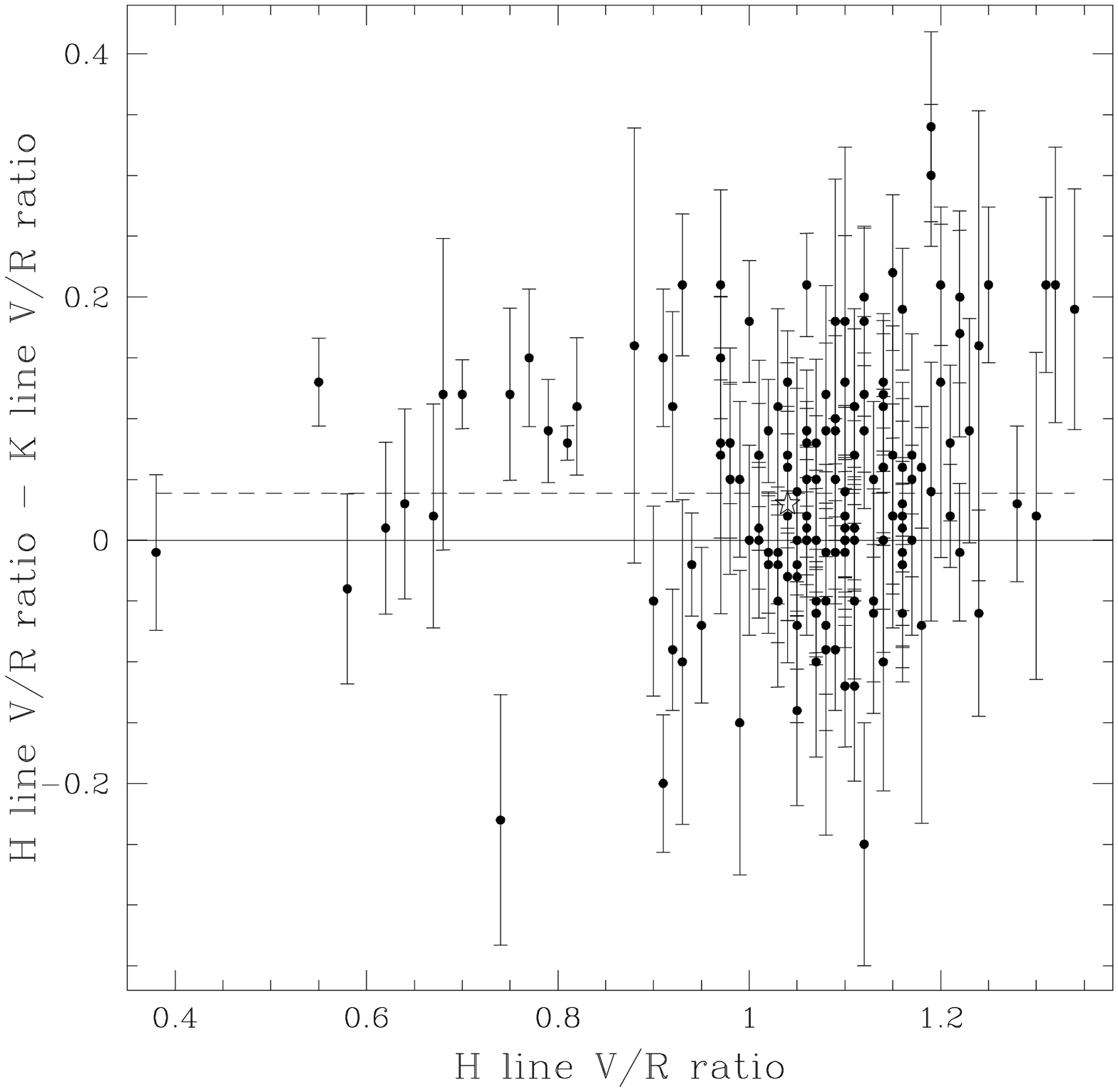}
\figcaption{The difference between the \ion{Ca}{2} H line and K line $V/R$ 
values plotted versus the measured H line $V/R$. The data are from Table 1, so certain stars may be represented by multiple points in this diagram. 
The single star symbol represents Arcturus. The dashed line shows the 
mean value of the $V/R$ parameter difference.
\label{fig5}}

\clearpage
\epsscale{0.8}
\plotone{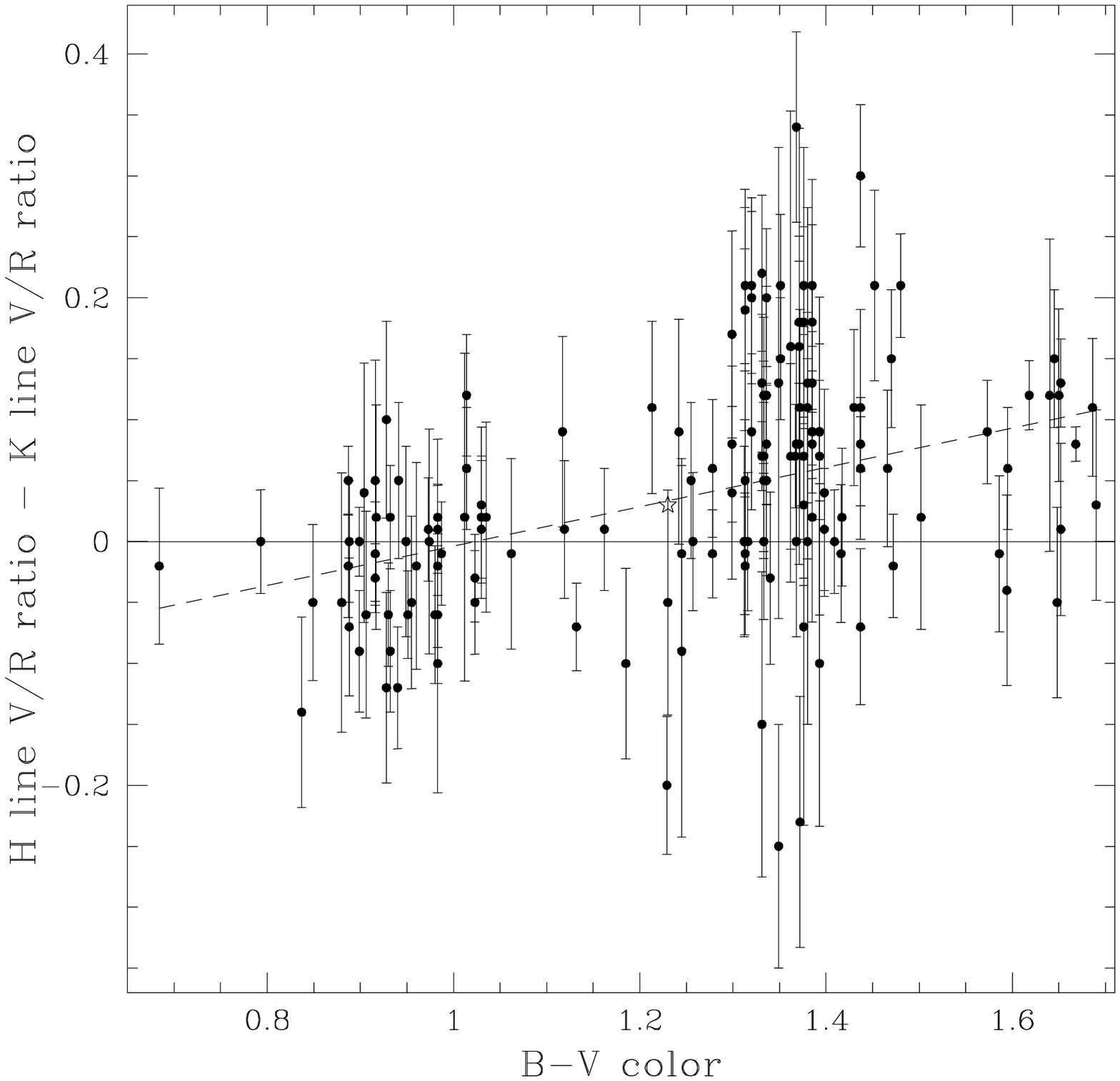}
\figcaption{Observed values of $(V/R)_H - (V/R)_K$ versus $B-V$ 
color. The data are from Table 1, so some stars show as multiple points 
in this diagram. The single star symbol represents Arcturus. The dashed line 
shows a least squares linear fit.
\label{fig6}}

\clearpage
\epsscale{0.8}
\plotone{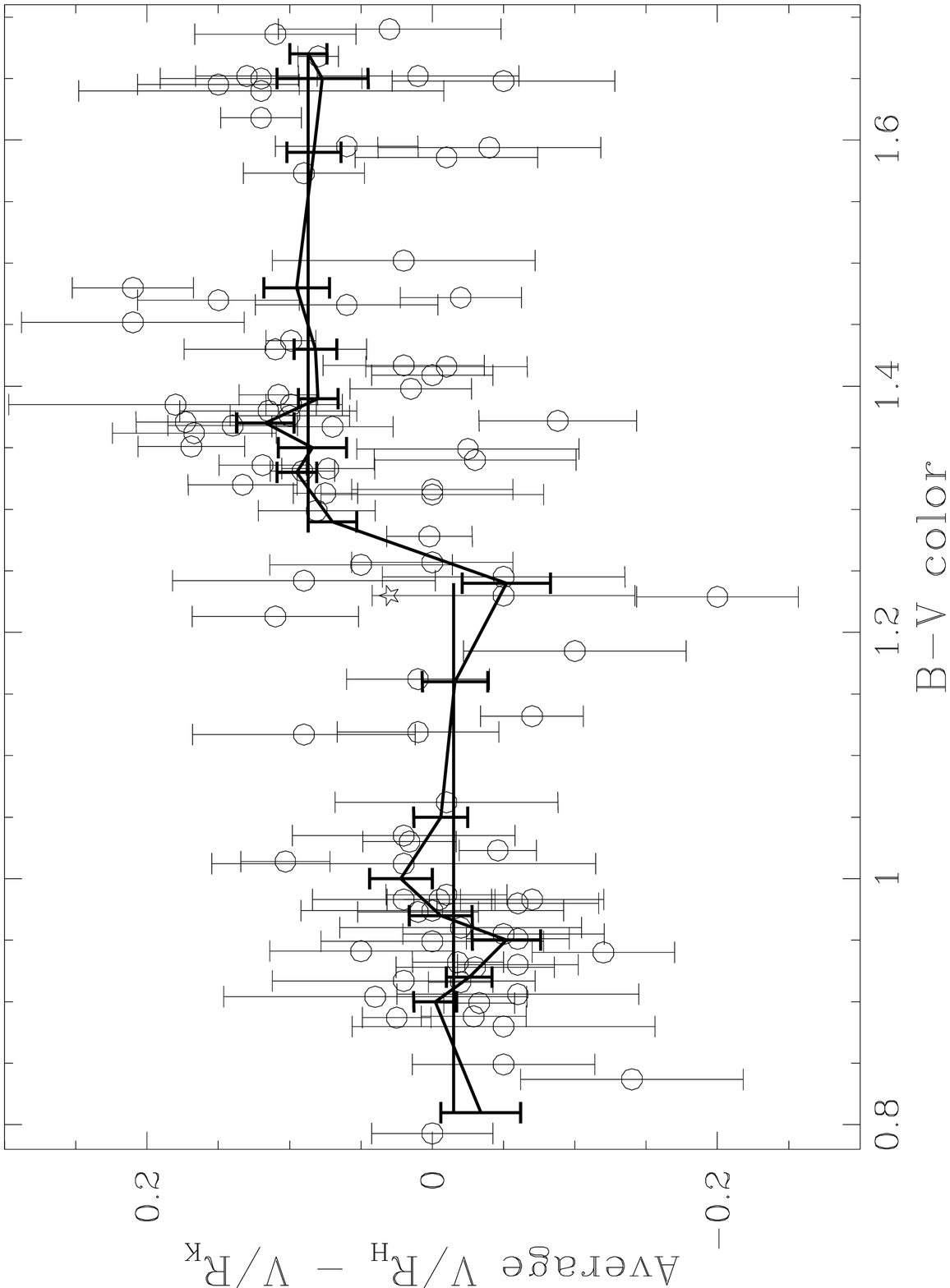}
\figcaption{The average $(V/R)_H - (V/R)_K$ value for each star 
listed in Table 2 versus $B-V$ color. A continuous line connects
the weighted mean of every 5 points binned according to $B-V$. This curve
shows a rather abrupt change at $B-V \sim 1.25$. The mean values on either
side of this ``break'' are shown by the horizontal lines.
\label{fig7}}

\clearpage
\epsscale{0.8}
\plotone{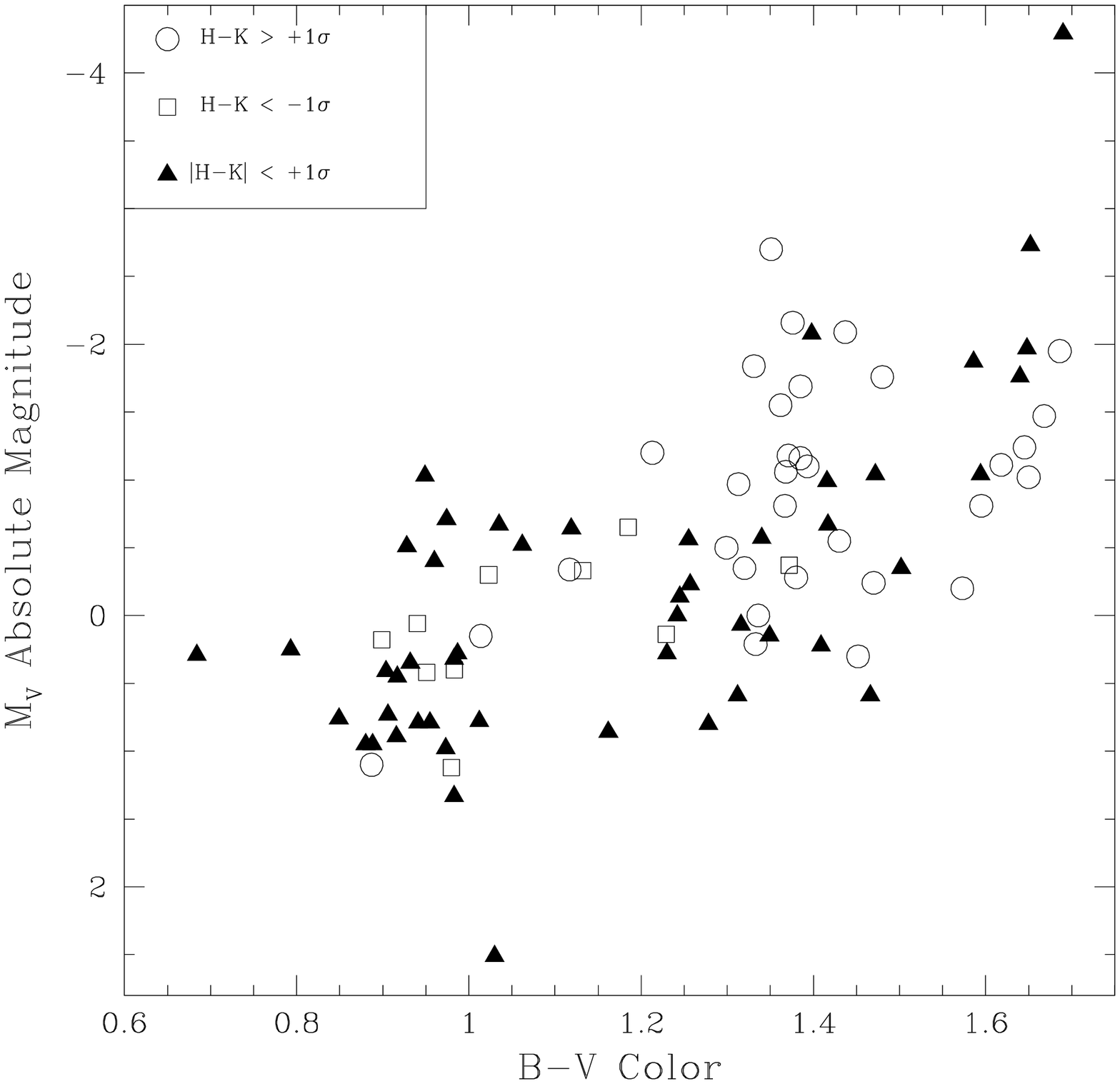}
\figcaption{The color magnitude diagram of the present sample of giants listed 
in Table 2. The open circles and open squares represent data points for which
the average $\langle (V/R)_H - (V/R)_K \rangle > 1\sigma$ or $< -1\sigma$ 
respectively. Filled triangles represent data points for which 
$\vert \langle (V/R)_H - (V/R)_K \rangle \vert < 1\sigma$.
\label{fig8}}  

\clearpage
\epsscale{0.8}
\plotone{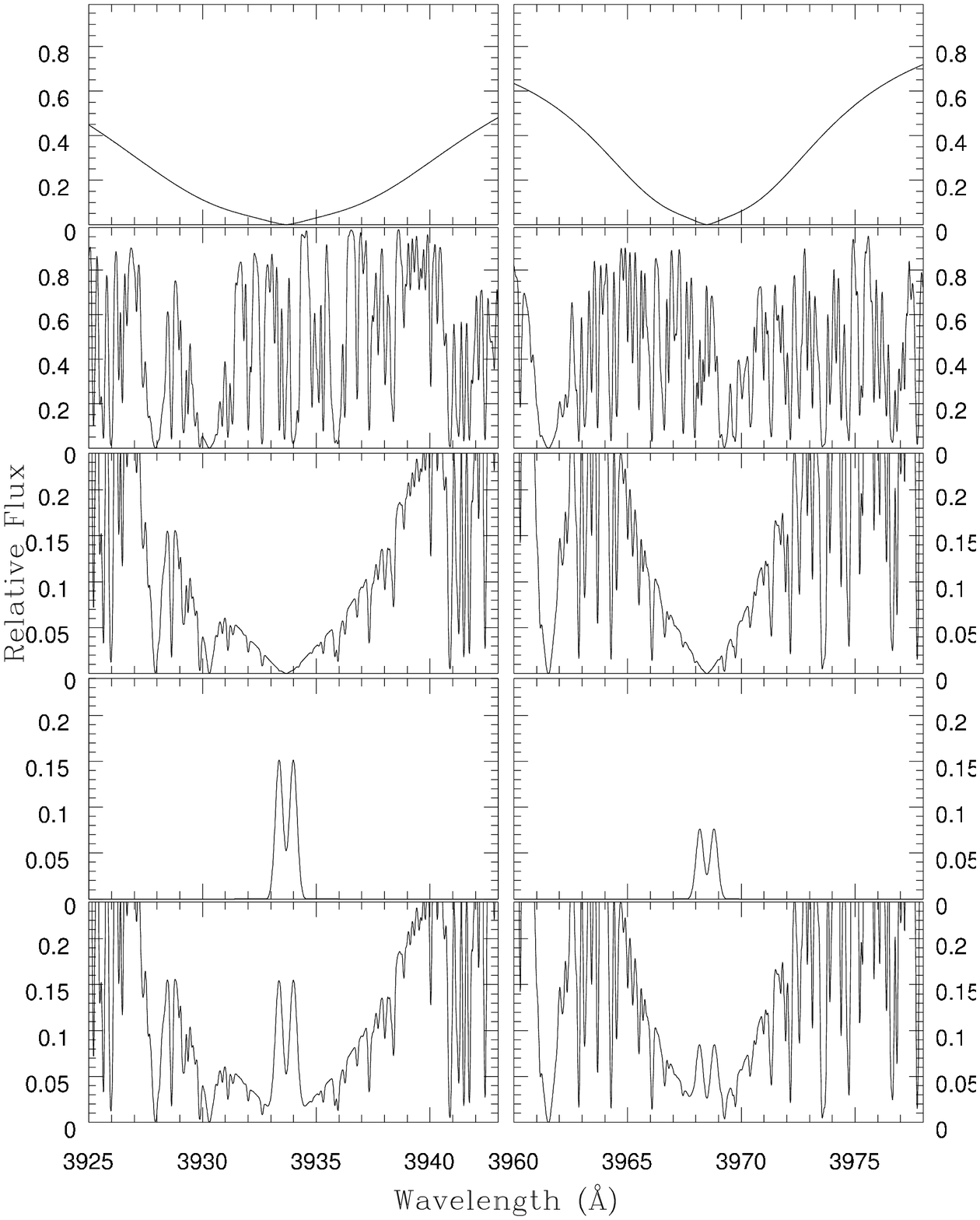}
\figcaption{Synthetic spectra at wavelengths encompassing the \ion{Ca}{2} H and
K lines for a star with $T_{\rm eff} = 4250$ K, $\log g = 1.75$, and 
[Fe/H] = --0.1. Horizontally paired panels correspond to the regions of 
the K (left) and H (right) lines. Running from top to bottom the panels show:
(i) the H and K photospheric line profiles alone,  
(ii) a photospheric spectrum based on a line list that excludes H and K, 
(iii) the full photospheric spectrum,
(iv) simulated chromospheric emission lines with a ratio of 2:1 in the K:H line
strengths, and 
(v) the summation of the complete synthetic
photospheric spectrum plus the simulated emission components.
\label{fig9}}

\clearpage
\epsscale{0.8}
\plotone{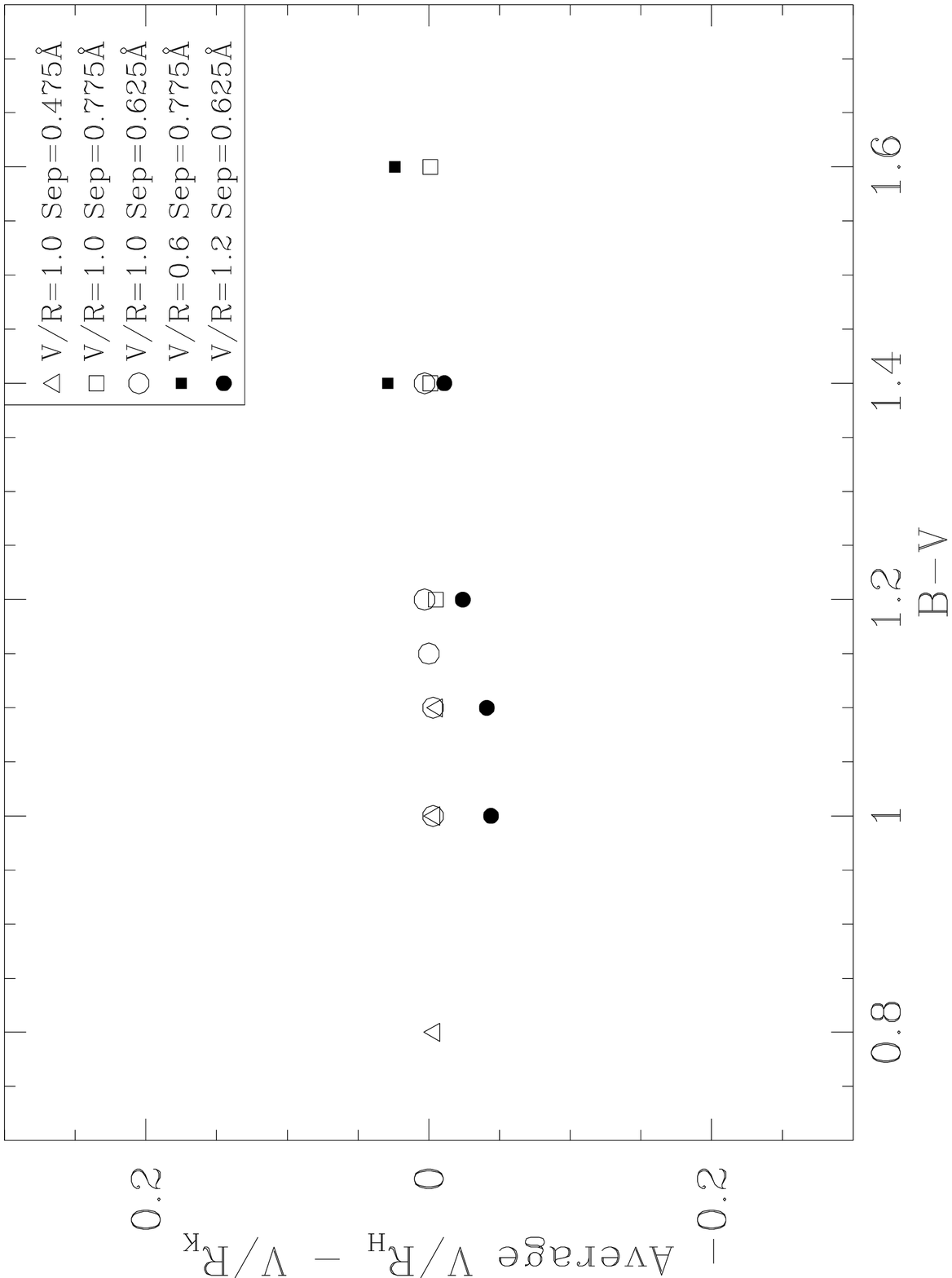}
\figcaption{The asymmetry difference $(V/R)_H - (V/R)_K$,
found after taking H and K emission lines of equal initial
$(V/R)$ and superimposing them on a photospheric spectrum,
versus the $(B-V)$ color of the photosphere. Symbols denote 
the morphology of the intrinsic H and K emission lines, 
which are parameterized by their initial $(V/R)$ values and the
peak-to-peak separation of the bimodal profiles. 
\label{fig10}}

\clearpage
\epsscale{0.8}
\plotone{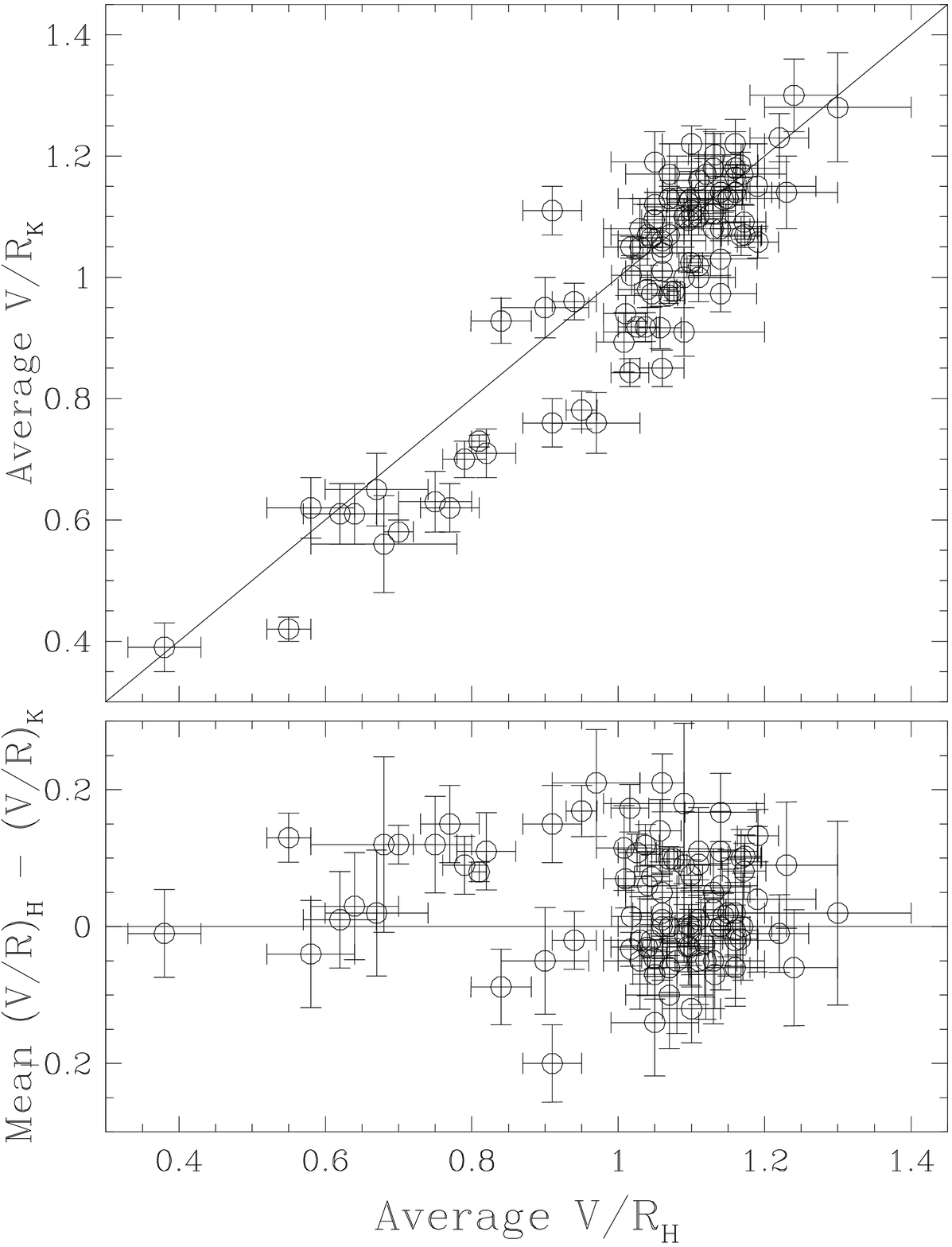}
\figcaption{The large upper panel shows the average value of $(V/R)_K$
versus the average $(V/R)_H$ for each star listed in Table 2. The lower 
panel plots the average value of the asymmetry difference 
$(V/R)_H - (V/R)_K$ versus average $(V/R)_H$.
The solid lines are the locus for $(V/R)_H = (V/R)_K$. 
\label{fig11}}

\clearpage
\epsscale{0.8}
\plotone{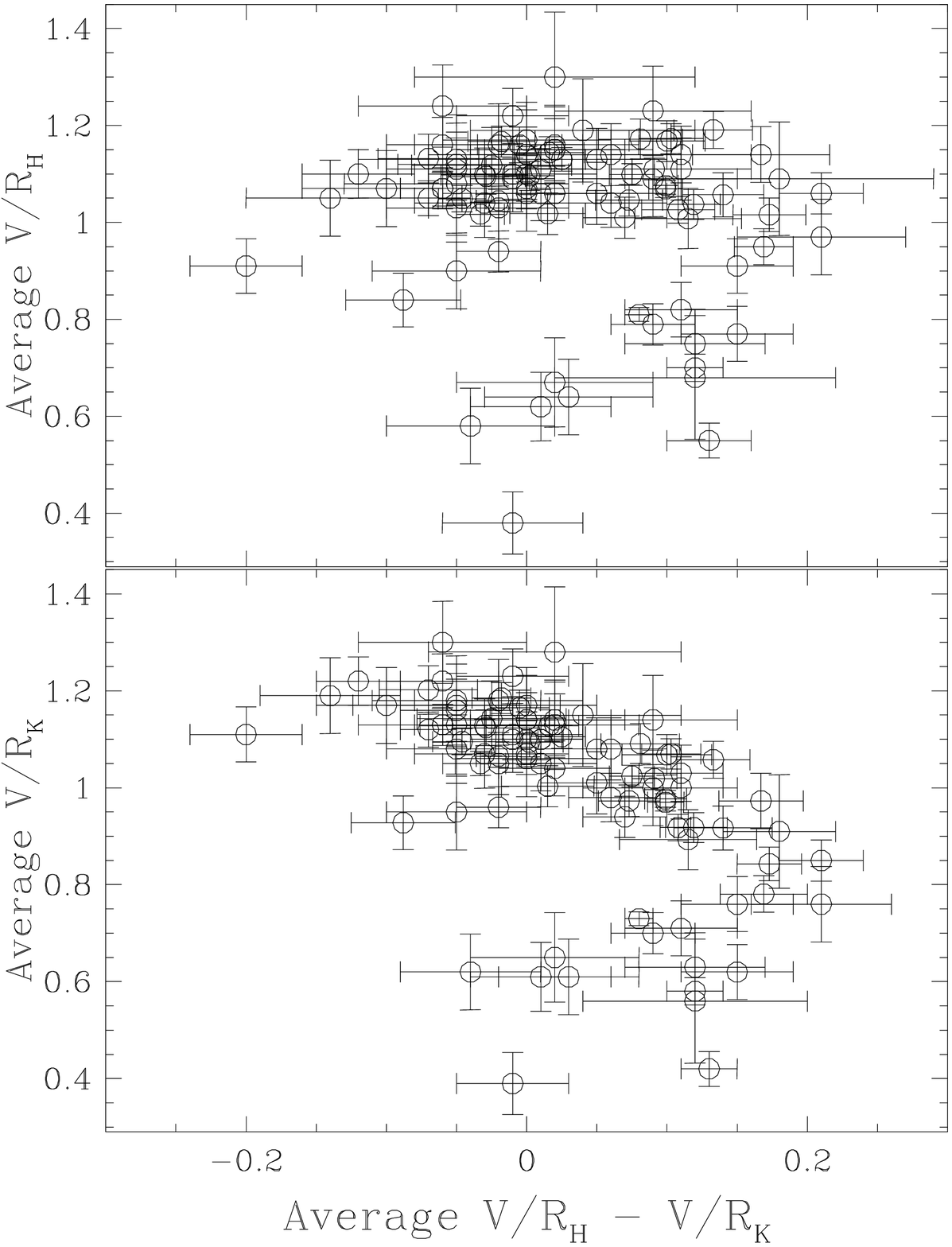}
\figcaption{The average value of $(V/R)_H$ and the average 
$(V/R)_K$ for each star in Table 2 plotted versus the average 
value of the asymmetry difference $(V/R)_H - (V/R)_K$. Unlike 
Figure 5, each star is represented by a single point in this figure.  
\label{fig12}}

\end{document}